\documentclass[a4paper,fleqn,review]{cas-dc}

\usepackage{amsmath,amssymb,amsfonts}
\usepackage{algorithmic}
\usepackage{graphicx}
\usepackage{textcomp}
\usepackage{comment,color}
\usepackage{enumitem}
\usepackage{epsfig}
\usepackage{caption}
\usepackage{verbatim,bm,multicol,blindtext}
\usepackage{epstopdf,array,mathtools,url,slashbox,subfigure,hyperref}
\usepackage[numbers]{natbib}
\usepackage[switch]{lineno}

\newcolumntype{L}{>{\centering\arraybackslash}m{3cm}}
\DeclarePairedDelimiter\norm{\lVert}{\rVert}%
\makeatletter
\let\oldabs\abs
\def\abs{\@ifstar{\oldabs}{\oldabs*}}
\let\oldnorm\norm
\def\norm{\@ifstar{\oldnorm}{\oldnorm*}}
\makeatother
\def\tsc#1{\csdef{#1}{\textsc{\lowercase{#1}}\xspace}}
\tsc{WGM}
\tsc{QE}
\tsc{EP}
\tsc{PMS}
\tsc{BEC}
\tsc{DE}

\begin{document}
\let\WriteBookmarks\relax
\def\floatpagepagefraction{1}
\def\textpagefraction{.001}
\shorttitle{Eikonal solution using PINNs}
\shortauthors{Waheed et~al.}

\title [mode=title]{{\color{black}PINNeik: Eikonal solution using physics-informed neural networks}}                    



\author[1]{Umair bin Waheed}[
                        orcid=0000-0002-5189-0694
                        ]

\cormark[1]
\ead{umair.waheed@kfupm.edu.sa}

\credit{Conceptualization, Methodology, Validation, Visualization, Software, Writing - original and draft}

\address[1]{Department of Geosciences, King Fahd University of Petroleum and Minerals, Dhahran 31261,  Saudi Arabia.}

\author[2]{Ehsan Haghighat}[]
\address[2]{Department of Civil Engineering, Massachusetts Institute of Technology, 
MA 02139, USA.}

\credit{Conceptualization, Methodology, Software, Writing - original and draft}

\author[3]{Tariq Alkhalifah}[]

\address[3]{Physical Sciences and Engineering Division, King
Abdullah University of Science and Technology, Thuwal 23955, Saudi Arabia.}

\credit{Supervision, Conceptualization, Validation, Writing - review and editing}

\author[3]{Chao Song}

\credit{Methodology, Visualization, Writing - review and editing}

\author[1]{Qi Hao}
\credit{Validation, Writing - review and editing}

\cortext[cor1]{Corresponding author}



\begin{abstract}
The eikonal equation is utilized across a wide spectrum of science and engineering disciplines. In seismology, it regulates seismic wave traveltimes needed for applications like source localization, imaging, and inversion. Several numerical algorithms have been developed over the years to solve the eikonal equation. However, these methods require considerable modifications to incorporate additional physics, such as anisotropy, and may even breakdown for certain complex forms of the eikonal equation, requiring approximation methods. Moreover, they suffer from computational bottleneck when repeated computations are needed for perturbations in the velocity model and/or the source location, particularly in large 3D models. Here, we propose an algorithm to solve the eikonal equation based on the emerging paradigm of physics-informed neural networks (PINNs). By minimizing a loss function formed by imposing the eikonal equation, we train a neural network to output traveltimes that are consistent with the underlying partial differential equation. 
We observe sufficiently high traveltime accuracy for most applications of interest. We also demonstrate how the proposed algorithm harnesses machine learning techniques like transfer learning and surrogate modeling to speed up traveltime computations for updated velocity models and source locations. Furthermore, we use a locally adaptive activation function and adaptive weighting of the terms in the loss function to improve convergence rate and solution accuracy. We also show the flexibility of the method in incorporating medium anisotropy and free-surface topography compared to conventional methods that require significant algorithmic modifications. These properties of the proposed PINN eikonal solver are highly desirable in obtaining a flexible and efficient forward modeling engine for seismological applications.
\end{abstract}


\begin{keywords}
eikonal equation \sep physics-informed neural networks \sep seismic modeling \sep traveltimes
\end{keywords}

\maketitle

\section{Introduction}

\label{sec:introduction}

The eikonal (from the Greek word $\mathrm{\epsilon \iota \kappa \omega \nu =}$~image) equation is a first-order non-linear partial differential equation (PDE) encountered in the wave propagation and geometric optics literature. It was first derived by Sir William Rowan Hamilton in the year 1831~\cite{masoliver2009classical}. The eikonal equation finds its roots in both wave propagation theory and geometric optics. In wave propagation, the eikonal equation can be derived from the first term of the Wentzel-Kramers-Brillouin (WKB) expansion of the wave equation~\cite{paris1969basic}, whereas in geometric optics, it can be derived using Huygen's principle~\cite{arnol2013mathematical}. 

Despite its origins in optics, the eikonal equation finds applications in many science and engineering problems. To name a few, in image processing, it is used to compute distance fields from one or more points~\cite{adalsteinsson1994fast}, inferring 3D surface shapes from intensity values in 2D images~\cite{rouy1992viscosity}, image denoising~\cite{malladi1996unified}, segmentation~\cite{alvino2007efficient}, and registration~\cite{cao2004registration}. In robotics, the eikonal equation is extensively used for optimal path planning and navigation, e.g., for domestic robots~\cite{ventura2014towards}, autonomous underwater vehicles~\cite{petres2007path}, and Mars Rovers~\cite{garrido2016path}. In computer graphics, the eikonal equation is used to compute geodesic distances for extracting shortest paths on discrete and parametric surfaces~\cite{spira2004efficient,raviv2011affine}. In semi-conductor manufacturing, the eikonal equation is used for etching, deposition, and lithography simulations~\cite{helmsen1996two,adalsteinsson1996level}. Furthermore, and of primary interest to us, the eikonal equation is routinely employed in seismology to compute traveltime fields needed for many applications, including statics and moveout correction~\cite{lawton1989computation}, traveltime tomography~\cite{guo2019first}, microseismic source localization~\cite{grechka2015relative}, and Kirchhoff migration~\cite{lambare20033d}.

The fast marching method (FMM) and the fast sweeping method (FSM) are the two most commonly used algorithms for solving the eikonal equation. FMM belongs to the family of algorithms which are also referred to as single-pass methods. The first such algorithm is attributed to John Tsitsiklis~\cite{tsitsiklis1995efficient}, who used a control-theoretic discretization of the eikonal equation and emulated Dijkstra-like shortest path algorithm. However, a few months later, a finite-difference approach, also based on Dijkstra-like ordering and updating was developed~\cite{sethian1996fast}. The FMM combines entropy satisfying upwind schemes for gradient approximations and a fast sorting mechanism to solve the eikonal equation in a single-pass.

The FSM, on the other hand, is a multi-pass algorithm that combines Gauss-Seidel iterations with alternating sweeping ordering to solve the eikonal equation~\cite{zhao2005fast}. The idea behind the algorithm is that the characteristics of the eikonal equation can be divided into a finite number of pieces and information propagating along each piece can be accounted for by one of the sweeping directions. Therefore, FSM converges in a finite number of iterations, irrespective of the grid size. 

Both FMM and FSM were initially proposed to solve the eikonal equation on rectangular grids. However, many different approaches have since been proposed, extending them to other discretizations and formulations. A detailed analysis and comparison of these fast methods can be found in~\cite{gomez2019fast}. 

On a different front, deep learning is fast emerging as a potential disruptive tool to tackle longstanding research problems across science and engineering disciplines~\cite{najafabadi2015deep}. Recent advances in the field of Scientific Machine Learning have demonstrated the largely untapped potential of deep learning for applications in scientific computing. The idea to use neural networks for solving PDEs has been around since the 1990s~\cite{lee1990neural,lagaris1998artificial}. However, recent advances in the theory of deep learning coupled with a massive increase in computational power and efficient graph-based implementation of new algorithms and automatic differentiation \cite{baydin2017automatic} have seen a resurgence of interest in using neural networks to approximate the solution of PDEs. 

This resurgence is confirmed by the advances made in the recent literature on scientific computing. For example,~\cite{ling2016reynolds} used a deep neural network (DNN) for modeling turbulence in fluid dynamics, while~\cite{han2018solving} proposed a deep learning algorithm to solve the non-linear Black–Scholes equation, the Hamilton–Jacobi–Bellman equation, and the Allen–Cahn equation. Similarly,~\cite{sirignano2018dgm} developed a mesh-free algorithm based on deep learning for efficiently solving high-dimensional PDEs. In addition,~\cite{tompson2017accelerating} used a convolutional neural network to speed up the solution to a sparse linear system required to obtain a numerical solution of the Navier-Stokes equation.

Recently, Raissi et al.~\cite{raissi2019physics} developed a deep learning framework for the solution and discovery of PDEs. The so-called physics-informed neural network (PINN) leverages the capabilities of DNNs as universal function approximators. In contrast with the conventional deep learning approaches, PINNs restrict the space of admissible solutions by enforcing the validity of the underlying PDE governing the actual physics of the problem. This is achieved by using a simple feed-forward network leveraging automatic differentiation (AD), also known as algorithmic differentiation. PINNs have already demonstrated success in solving a wide range of non-linear PDEs, including Burgers, Schr\"{o}dinger, Navier-Stokes, and Allen-Cahn equations~\cite{raissi2019physics}. Moreover, PINNs have also been successfully applied to problems arising in geosciences~\cite{xu2019physics,karimpouli2020physics,song2021solving,bai2021accelerating,waheed2021pinntomo}.

In this paper, we propose a paradigm shift from conventional numerical algorithms to solve the eikonal equation. Using a loss function defined by the underlying PDE, we train a DNN to yield the solution of the eikonal equation. To mitigate point-source singularity, we use the factored eikonal equation. Through tests on benchmark synthetic models, we study the accuracy properties of the proposed solver. We also explore how machine learning techniques like transfer learning and surrogate modeling can potentially speed up repeated traveltime computations with updated velocity models and/or source locations. We also demonstrate the flexibility of the proposed scheme in incorporating additional physics and surface topography into the eikonal solution.

The main contributions of this paper are as follows: (1)~We propose a novel algorithm to solve the eikonal equation based on neural networks, which predicts functional solutions by setting the underlying PDE as a loss function to optimize the network's parameters. The proposed algorithm achieves sufficiently high accuracy on models of practical interest. (2) Through the use of transfer learning, we show how repeated traveltime computations can be done efficiently. On the contrary, conventional algorithms like fast marching and fast sweeping require the same computational effort even for small perturbations in the velocity model or source location. (3) We demonstrate that by constructing surrogate models with respect to the source location, the computations can be sped up dramatically as only a single evaluation of the trained neural network is needed for perturbations in the source location. Such a model can also be effectively used for sensitivity analysis. (4) We demonstrate the flexibility of the proposed approach in incorporating additional physics by simply updating the loss function and the fact that no special treatment is needed to accurately account for surface topography or any irregularly shaped domain.

The rest of the paper is organized as follows. We begin by describing the theoretical underpinnings of the algorithm. 
Then, we present numerical tests probing into the accuracy of the proposed framework on synthetic velocity models. We also explore the applicability of transfer learning and surrogate modeling to efficiently solve the eikonal equation. Next, we discuss the strengths and limitations of the approach, including implications of this work on the field of numerical eikonal solvers. This is followed by some concluding remarks.

\section{Theory}
In this section, we first introduce the eikonal equation and the factorization idea. This is followed by a brief overview of deep neural networks and their capabilities as function approximators. Next, we briefly explain the concept of automatic differentiation. Finally, putting these pieces together, we present the proposed algorithm for solving the eikonal equation.

\subsection{Eikonal equation}
The eikonal equation is a non-linear, first-order, hyperbolic PDE of the form:

\begin{equation}
\begin{aligned}
    |\nabla T (\mathbf{x})|^2 & = \frac{1}{v^2(\mathbf{x})}, \qquad \forall \, \mathbf{x}\, \in \, \Omega, \\
   T(\mathbf{x_s}) & = 0,
\label{eq:eikonal}
\end{aligned}
\end{equation}

\noindent where $\Omega$ is a domain in $\mathbb{R}^d$ with $d$ as the space dimension, $T(\mathbf{x})$ is the traveltime or Euclidean distance to any point $\mathbf{x}$ from the point-source $\mathbf{x_s}$, $v(\mathbf{x})$ is the velocity defined on $\Omega$, and $\nabla$ denotes the spatial differential operator. Equation~\eqref{eq:eikonal} simply means the gradient of the arrival time surface is inversely proportional to the speed of the wavefront. This is also commonly known as the isotropic eikonal equation as the velocity is not a function of the wave propagation direction $(\nabla T/|\nabla T|)$. 

To avoid the singularity due to the point-source~\cite{qian2002adaptive}, we consider the factored eikonal equation~\cite{fomel2009fast}. The factorization approach relies on factoring the unknown traveltime ($T (\mathbf{x})$) into
two functions. One of the functions is specified analytically, such that the other function is smooth in the source neighborhood. Specifically, we consider multiplicative factorization, i.e.,

\begin{equation}
    T(\mathbf{x}) = T_0(\mathbf{x}) \, \tau(\mathbf{x}),
    \label{eq:factorization}
\end{equation}

\noindent where $T_0(\mathbf{x})$ is known and $\tau(\mathbf{x})$ is the unknown function. Plugging equation~\ref{eq:factorization} in equation~\ref{eq:eikonal}, we get the factored eikonal equation:

\begin{equation}
\begin{aligned}
    T_0^2\,|\nabla \tau|^2 + \tau^2\,|\nabla T_0|^2 & + 2 \, T_0 \, \tau \, (\nabla T_0 . \nabla \tau)  = \frac{1}{v^2(\mathbf{x})},\\
   \tau(\mathbf{x_s}) & = 1.
\label{eq:fac_eikonal}
\end{aligned}
\end{equation}

\noindent The known factor $T_0$ is computed analytically using the expression:

\begin{equation}
    T_0(\mathbf{x}) = \frac{|\mathbf{x} - \mathbf{x_s}|}{v(\mathbf{x_s})},
    \label{eq:knownsol}
\end{equation}

\noindent where $v(\mathbf{x_s})$ is the velocity at the source location.

\subsection{Deep feed-forward neural networks}
\begin{figure}
\centering
  \includegraphics[width=0.46\textwidth]{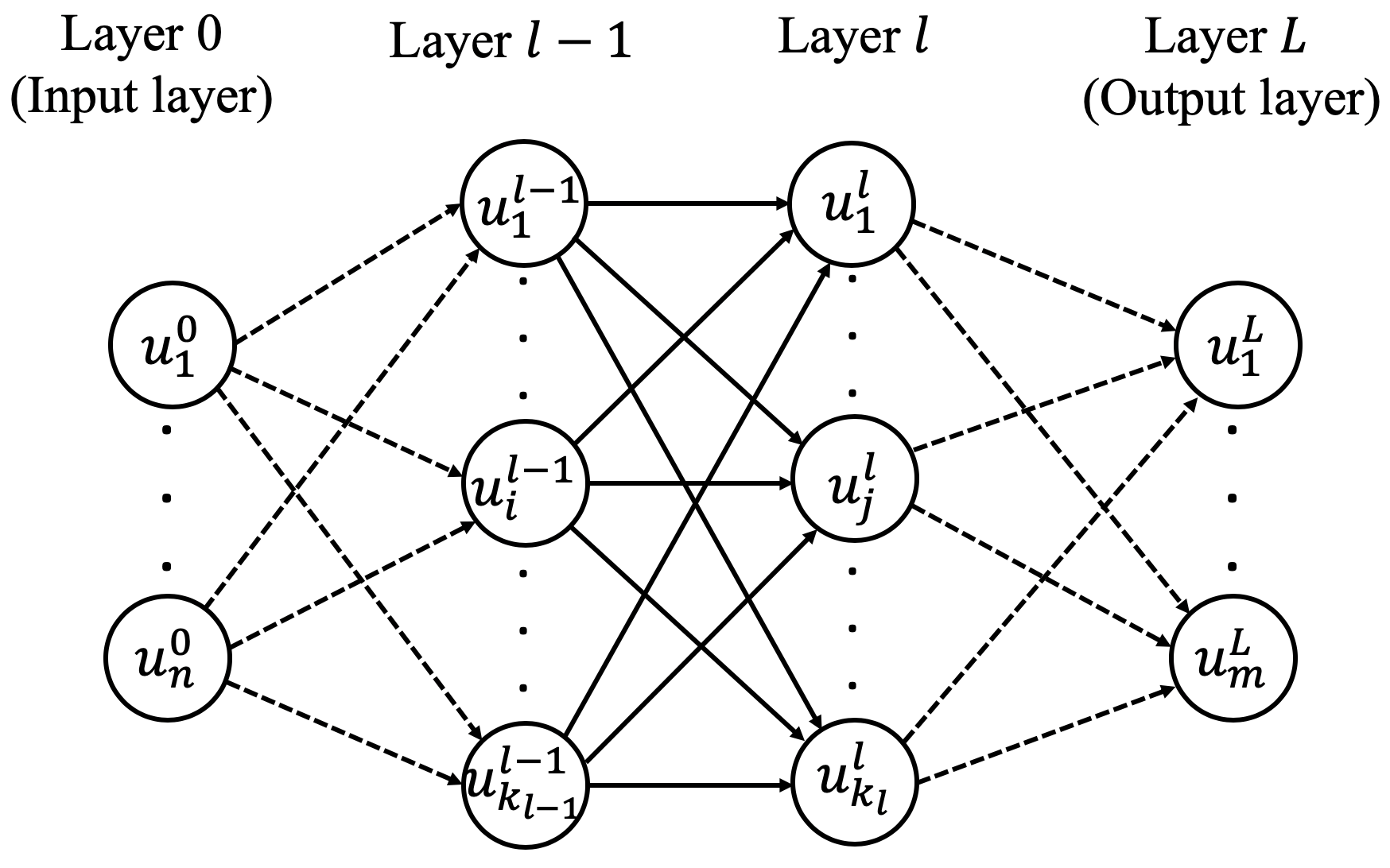}\\
    \caption{Schematic representation of a feed-forward neural network with $L-1$ hidden layers.}
  \label{fig:ann_schematic}
\end{figure}

A feed-forward neural network is a set of neurons organized in layers in which evaluations are performed sequentially through the layers. It can be seen as a computational graph having an input layer, an output layer, and an arbitrary number of hidden layers. In a fully connected neural network, neurons in adjacent layers are connected with each other but neurons within a single layer share no connection. 

Thanks to the universal approximation theorem, a neural network with $n$ neurons in the input layer and $m$ neurons in the output layer can be used to represent a multi-dimensional function $u: \mathbb{R}^n \rightarrow \mathbb{R}^m$ ~\cite{hornik1989multilayer},
as shown in Figure~\ref{fig:ann_schematic}. For illustration, we consider a network of $L+1$ layers starting with input layer 0, the output layer $L$, and $L-1$ hidden layers. The number of neurons in each layer is denoted as $k_0=n, k_1, \cdots, k_L=m$. Each connection between the $i$-th neuron in layer $l-1$ and $j$-th neuron in layer $l$ has a weight $w_{ji}^l$ associated with it. Moreover, for each neuron in layer $l$, we have an associated bias term $b_i, i = 1, \cdots, k_l$. Each neuron represents a mathematical operation, whereby it takes a weighted sum of its inputs plus a bias term and passes it through an activation function. The output from the $k$-th neuron in layer $l$ is given as~\cite{bishop2006pattern}:

\begin{equation}
    u_k^l = \sigma \left( \sum_{j=1}^{k_l-1} w_{kj}^l \; u_j^{l-1} + b_k^l \right),
\label{eq:neuron}
\end{equation}

\noindent where $\sigma()$ represents the activation function. Commonly used activation functions are the logistic sigmoid, the hyperbolic tangent, and the rectified linear unit~\cite{sibi2013analysis}. By dropping the subscripts, we can write equation~\eqref{eq:neuron} compactly in the vectorial form:

\begin{equation}
    \mathbf{u}^l = \sigma \left( \mathbf{W}^l \mathbf{u}^{l-1} + \mathbf{b}^l  \right),
\label{eq:neuronvec}
\end{equation}

\noindent where $\mathbf{W}^l$ is the matrix of weights corresponding to connections between layers $l-1$ and $l$, $\mathbf{u}^l$ and $\mathbf{b}^l$ are vectors given by $u_k^l$ and $b_k^l$, respectively, and the activation function is applied element-wise. Computational frameworks, such as $\texttt{Tensorflow}$~\cite{tensorflow2015-whitepaper}, can be used to efficiently evaluate data flow graphs like the one given in equation~\eqref{eq:neuronvec} efficiently using parallel execution. The input values can be defined as tensors (multi-dimensional arrays) and the computation of the outputs is vectorized and distributed across the available computational resources for efficient evaluation.

\subsection{Approximation property of neural networks}

Neural networks are well-known for their strong representational power. It has been shown that a neural network with a single hidden layer and a finite number of neurons can be used to represent any bounded continuous function to any desired accuracy. This is also known as the universal approximation theorem~\cite{cybenko1989approximation,hornik1989multilayer}. It was later shown that by using a non-linear activation function and a deep network, the total number of neurons can be significantly reduced~\cite{lu2017expressive}. Therefore, we seek a trained deep neural network (DNN) that could represent the mapping between the input ($\mathbf{x}$) and the output ($\tau(\mathbf{x})$) of the factored eikonal equation for a given velocity model ($v(\mathbf{x})$).

It is worth noting that while neural networks are, in theory, capable of representing very complex functions compactly, finding the actual parameters (weights and biases) needed to solve a given PDE can be quite challenging.

\subsection{Automatic differentiation}

Solving a PDE using PINNs requires derivatives of the network's output with respect to the inputs. There are four possible ways to compute derivatives~\cite{baydin2017automatic,margossian2019review}: (1)~hand-coded analytical derivatives, (2)~symbolic differentiation, (3)~numerical approximation such as finite-difference, and (4)~automatic differentiation (AD). 

Manually working out the derivatives may be exact, but they are not automated, and thus, impractical. Symbolic differentiation is also exact, but it is memory intensive and prohibitively slow as one could end up with exponentially large expressions to evaluate. While numerical differentiation is easy to implement, it can be highly inaccurate due to round-off errors. On the contrary, AD uses exact expressions with floating-point values instead of symbolic strings and it involves no approximation error, resulting in accurate evaluation of derivatives at machine precision. However, an efficient implementation of AD can be non-trivial. Fortunately, many existing computational frameworks such as $\texttt{Tensorflow}$~\cite{tensorflow2015-whitepaper} and $\texttt{PyTorch}$~\cite{adam2017automatic} have made available efficiently implemented AD libraries. In fact, in deep learning, backpropagation~\cite{rumelhart1986learning}, a generalized technique of AD, has been the mainstay for training neural networks. 

To understand how AD works, consider a simple fully-connected neural network with two inputs $(x_1,x_2)$, one output $(y)$, and one neuron in the hidden layer. Let us assume the network's weights and biases are assigned such that:

\begin{equation}
    \begin{aligned}
    \nu & = 2x_1 + 3x_2 - 1,\\
    h & = \sigma(\nu) = \frac{1}{1+e^{-\nu}},\\
    y & = 5h + 2,
    \end{aligned}
\label{eq:adexample}
\end{equation}

where $h$ represents the output from the neuron in the hidden layer computed by applying the sigmoid function $(\sigma)$ on the weighted sum of the inputs $(\nu)$. 

To illustrate the idea, let us say we are interested in computing partial derivatives $\frac{\partial y}{x_1}$ and $\frac{\partial y}{x_2}$ at $(x_1,x_2) = (1,-1)$. AD requires one forward pass and backward pass through the network to compute these derivatives as detailed in Table~\ref{tbl:table1}. To compute high-order derivatives, AD can be applied recursively through the network in the same manner. For a deeper understanding of AD, we refer the interested reader to~\cite{elliott2018simple}.

\begin{table}[h!]
\centering
\begin{tabular}{|p{32mm}|p{45mm}|}
  \hline
  \begin{center}
  \vspace{-0.25cm}
      {\large Forward pass} \vspace{-0.25cm}
  \end{center} &     \begin{center} \vspace{-0.25cm}
      {\large Reverse pass} \vspace{-0.25cm}
  \end{center}\\
  \hline \hline
  \vspace{-0.3cm}
  \begin{equation*} \begin{aligned}
    x_1 &= 1\\
    x_2 & = -1
    \end{aligned} \end{equation*} \vspace{-0.3cm}
 &   \vspace{-0.3cm} \begin{equation*} \begin{aligned}
    \frac{\partial y}{\partial y} = 1
    \end{aligned} \end{equation*} \vspace{-0.3cm} \\
  \hline
  \vspace{-0.2cm}
  \begin{equation*} \begin{aligned}
    \nu & = 2x_1 + 3x_2 - 1 = -2\\
    h & = \frac{1}{1+e^{-\nu}} = 0.119
    \end{aligned} \end{equation*}
 &   \vspace{-0.4cm} \begin{equation*} \begin{aligned}
    \frac{\partial y}{\partial h} & = \frac{\partial (5h+2)}{\partial h} = 5 \\
    \frac{\partial y}{\partial \nu} & = \frac{\partial y}{\partial h} . \frac{\partial h}{\partial \nu} = 5 \times \frac{e^{-\nu}}{(1+e^{-\nu})^2} \\
    & = 0.525
    \end{aligned} \end{equation*} \vspace{-0.2cm} \\
  \hline
  \vspace{-0.2cm}
  \begin{equation*} 
  y = 5h + 2 = 2.596 
  \end{equation*}
 &   \vspace{-0.4cm} \begin{equation*} \begin{aligned}
    \frac{\partial y}{\partial x_1} & = \frac{\partial y}{\partial \nu}. \frac{\partial \nu}{\partial x_1} = \frac{\partial y}{\partial \nu} \times 2 = 1.050\\
    \frac{\partial y}{\partial x_2} & = \frac{\partial y}{\partial \nu}. \frac{\partial \nu}{\partial x_2} = \frac{\partial y}{\partial \nu} \times 3 = 1.575
    \end{aligned} \end{equation*} \vspace{-0.1cm} \\
  \hline
\end{tabular}
 \caption{Example of forward and reverse pass computations needed by AD to compute partial derivates of the output with respect to the inputs at $(x_1,x_2) = (1,-1)$ for the expressions given in equation~\eqref{eq:adexample}.}
 \label{tbl:table1}
\end{table}

\subsection{Solving the eikonal equation}

\begin{figure*}
\begin{center}
\includegraphics[width=0.85\textwidth]{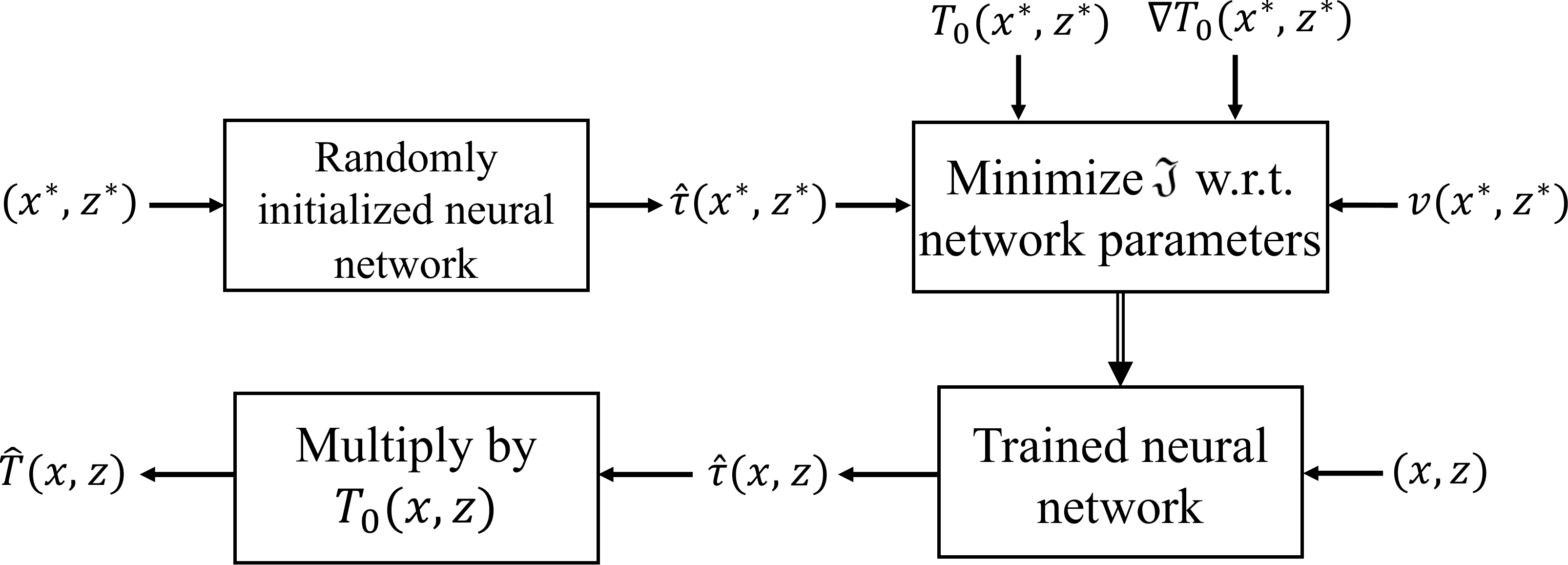}
\end{center}
\caption{
A workflow for the proposed eikonal solver: A randomly initialized neural network is trained on a set of randomly selected collocation points $(x^{*},z^{*})$ in the model space with given velocity $v(x^{*},z^{*})$ and the known traveltime function $T_0(x^{*},z^{*})$ and its spatial derivative $\nabla T_0(x^{*},z^{*})$ to minimize the loss function given in equation~\ref{eq:loss_mse}. Once the network is trained, it is evaluated on a regular grid of points $(x,z)$ to yield an estimate of the unknown traveltime field $\hat{\tau}$, which is then multiplied with the known traveltime part $T_0$ to yield the estimated eikonal solution $\hat{T}$.
}%
\label{fig:flowchart}
\end{figure*}

To solve the eikonal equation~\eqref{eq:eikonal}, we leverage the capabilities of neural networks as function approximators and define a loss function that minimizes the residual of the factored eikonal equation at a chosen set of training (collocation) points. This is achieved with (i) a DNN approximation of the unknown traveltime field variable $\tau(\mathbf{x})$; (ii) a loss function incorporating the eikonal equation and sampled on a collocation grid; (iii) a differentiation algorithm, i.e., AD in this case, to evaluate partial derivatives of $\tau(\mathbf{x})$ with respect to the spatial coordinates; and (iv) an optimizer to minimize the loss function by updating the network parameters.

\begin{figure}[ht!]
\begin{center}
\includegraphics[width=0.42\textwidth]{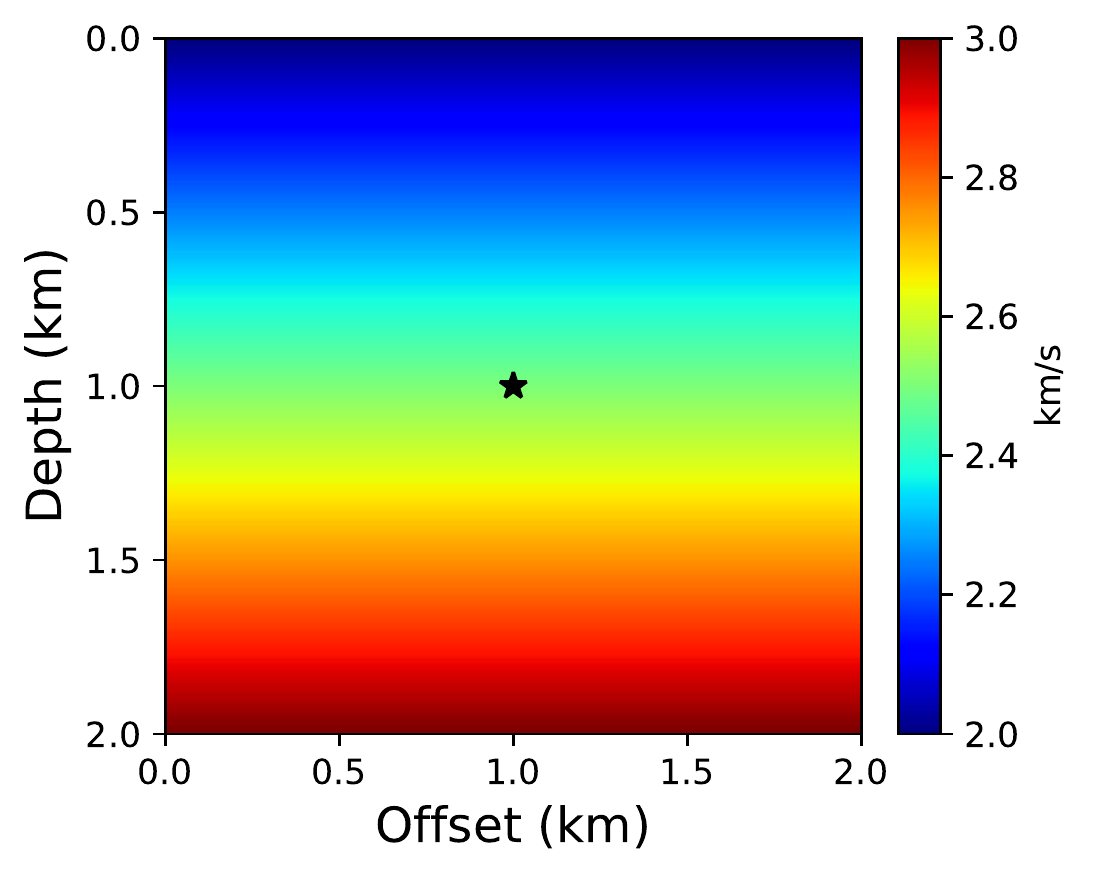}
\end{center}
\caption{
A velocity model with a constant velocity gradient of 0.5~$\text{s}^{-1}$ in the vertical direction. The velocity at zero depth is equal to 2~km/s and it increases linearly to 3~km/s at a depth of 2~km. Black star indicates the point-source location used for the tests.
}%
\label{fig:vofz_velmodel}
\end{figure}

To illustrate the idea, let us consider a two-dimensional domain $\Omega \in \mathbb{R}^2$ where $\mathbf{x} = (x,z) \in [0, 2]$, as shown in Figure~\ref{fig:vofz_velmodel}. A source term is considered at $\mathbf{x_s} = (x_s, z_s)$, where $\tau(\mathbf{x_s})=1$. The unknown traveltime factor $\tau(\mathbf{x})$ is approximated by a multilayer DNN $\mathcal{N}_\tau$, i.e., $\tau(\mathbf{x}) \approx \hat{\tau}(\mathbf{x}) = \mathcal{N}_\tau(\mathbf{x}; \boldsymbol{\theta})$, where $\mathbf{x} = (x,z)$ are network inputs, $\hat{\tau}$ is the network output, and  $\boldsymbol{\theta}$ represents the set of all trainable parameters of the network. 

The loss function can now be constructed using a mean-squared-error (MSE) norm as: 

\begin{equation}
\begin{aligned}
\mathfrak{J} & = \frac{1}{N_I}\sum_{\mathbf{x}^* \in I} \norm{\mathcal{L} }^2 + \frac{1}{N_I}\sum_{\mathbf{x}^* \in I} \norm{\mathcal{H}(-\hat{\tau})|\hat{\tau}|}^2 \\  
&+ \norm{\hat{\tau}(\mathbf{x_s}) - 1}^2,
\end{aligned}
\label{eq:loss_mse}
\end{equation}

\noindent where

\begin{equation}
\begin{aligned}
    \mathcal{L} &= T_0^2\,|\nabla \hat{\tau}|^2 + \hat{\tau}^2\,|\nabla T_0|^2\\
    & + 2 \, T_0 \, \hat{\tau} \, (\nabla T_0 . \nabla \hat{\tau})  - \frac{1}{v^2(\mathbf{x})},
\end{aligned}
\end{equation}
forms the residual of the factored eikonal equation. 

The first term on the right side of equation~\ref{eq:loss_mse} imposes the validity of the factored eikonal equation on a given set of training points $\mathbf{x}^*~\in~I$, with $N_I$ as the number of sampling points. The second term forces the solution $\hat{\tau}$ to be positive by penalizing negative solutions using the Heaviside function $\mathcal{H}()$. The last term requires the solution to be unity at the point-source $\mathbf{x_s} = (x_s,z_s)$. 

Network parameters $\boldsymbol{\theta}$ are then identified by minimizing the loss function \eqref{eq:loss_mse} on a set of sampling (training) points $\mathbf{x}^* \in I$, i.e.,

\begin{equation}
\arg\min_{\boldsymbol{\theta}} \mathfrak{J}(\mathbf{x}^*; \boldsymbol{\theta}).
\label{eq:optimization}
\end{equation}

Once the DNN is trained, we evaluate the network on a set of regular grid-points in the computational domain to obtain the unknown traveltime part. The final traveltime solution is obtained by multiplying it with the known traveltime part, i.e.,

\begin{equation}
    \hat{T}(\mathbf{x}) = T_0(\mathbf{x}) \, \hat{\tau}(\mathbf{x}).
\end{equation}

\noindent A pictorial description of the proposed algorithm is shown in Figure~\ref{fig:flowchart}. It should be highlighted that the computations of derivatives on the velocity model boundaries using AD is straightforward and do not need any special treatment.

It is also worth emphasizing that the proposed approach is different from traditional (or non-physics constrained) deep learning techniques. The training of the network here refers to the tuning of weights and biases of the network such that the resulting solution minimizes the loss function $\mathfrak{J}$ on a given set of training points. The training set here refers to the collocation points, usually chosen randomly, from within the computational domain. The number of collocation points needed to obtain a sufficiently accurate solution increases with the heterogeneity of the velocity model. 

Contrary to supervised learning applications, the network here learns without any labeled set. To understand this point, consider a randomly initialized network, which will output a certain value $\hat{\tau}_{i,j}$ for each point $(i,j)$ in the training set. These output values will be used to calculate the residual using equation~\ref{eq:loss_mse}. Based on this residual, the network adjusts its weights and biases, allowing it to produce $\hat{\tau}$ that adheres to the underlying factored eikonal equation~\eqref{eq:fac_eikonal}. 

\section{Numerical Tests}

In this section, we test the proposed PINN eikonal solver for computing traveltimes emanating from a point-source. We consider several velocity models, including a highly heterogeneous portion from the Marmousi model. We also include a model with irregular topography and anisotropy to demonstrate the flexibility of the proposed method compared to conventional algorithms.

For each example presented below, we use a neural network with 10 hidden layers containing 20 neurons in each layer and minimize the neural network's loss function using full-batch optimization. The input layer consists of two neurons, one for each spatial coordinate $(x,z)$, and the output layer has a single neuron to provide the estimated traveltime factor $\hat{\tau}(x,z)$. The network architecture is chosen using some initial tests and kept fixed for the entire study to avoid the need for architecture tuning for each new velocity model.

The activation function also plays an important role in the optimization of the network. We use a locally adaptive inverse tangent function for all hidden layers except the final layer, which has a linear activation function. Locally adaptive activation functions have recently shown better learning capabilities than the traditional or fixed activation functions in achieving higher convergence rate and solution accuracy~\cite{jagtap2020locally}. Using a scalable parameter in the activation function for each neuron changes the slope of the activation function and, therefore, alters the loss landscape of the network, yielding improved performance. 

Moreover, recent investigations into the PINN model have shown that the optimization process suffers from the discrepancy of the convergence rate in the different components of the loss function~\cite{wang2020and}. Using the statistics of the back-propagated gradient, adaptive weights could be assigned to different terms in the loss function, balancing the magnitude of back-propagated gradients. We adopt this strategy to adaptively assign weights to each term in the loss function for improved convergence. For more information on the adaptive weighting strategy, we refer the interested reader to~\cite{wang2020understanding,wang2020and}. 


The PINN framework is implemented using the \texttt{SciANN} package~\cite{haghighat2021sciann} -- a high level $\texttt{Tensorflow}$ wrapper for scientific computations. For comparison, we use the first-order finite-difference fast sweeping solution, which is routinely used for traveltime computations in seismological applications. 

First, we consider a $2 \times 2$ km$^2$ model with vertically varying velocity. The velocity at zero depth is 2~km/s and it increases linearly with a gradient of 0.5 s$^{-1}$. We consider the point-source to be located at $(1~\text{km},1~\text{km})$. The model is shown in Figure~\ref{fig:vofz_velmodel} with the black star depicting the point-source location. The model is discretized on a 101~$\times$~101 grid with a grid spacing of 20~m along both axes. For a model with a constant velocity gradient, the analytical traveltime solution is given as~\cite{slotnick1959lessons}:

\begin{equation}
    T(\mathbf{x}) = \frac{1}{\sqrt{g_v^2 + g_h^2}} \, \text{arccosh}\left(1 + \frac{\left(g_v^2 + g_h^2\right) |\mathbf{x} - \mathbf{x_s}|^2}{2 \, v(\mathbf{x}) \, v(\mathbf{x_s})}\right),
    \label{eq:analytical}
\end{equation}

where $T(\mathbf{x})$ is the traveltime value at some grid point $\mathbf{x}$ from a point-source located at $\mathbf{x_s}$. Likewise, $v(\mathbf{x})$ is the velocity at the grid-point $\mathbf{x}$ and $v(\mathbf{x_s})$ is the velocity at the point-source location. The velocity gradients along the vertical and horizontal dimensions are denoted by $g_v$ and $g_h$, respectively. Therefore, for the model in Figure~\ref{fig:vofz_velmodel}, $g_v = 0.5~\text{s}^{-1}$, $g_h = 0~\text{s}^{-1}$, $\mathbf{x_s} = (1~\text{km},1~\text{km})$, and $v(\mathbf{x_s}) = 2.5~\text{km/s}$.

\begin{figure}[ht!]
\begin{center}
\includegraphics[width=0.47\textwidth]{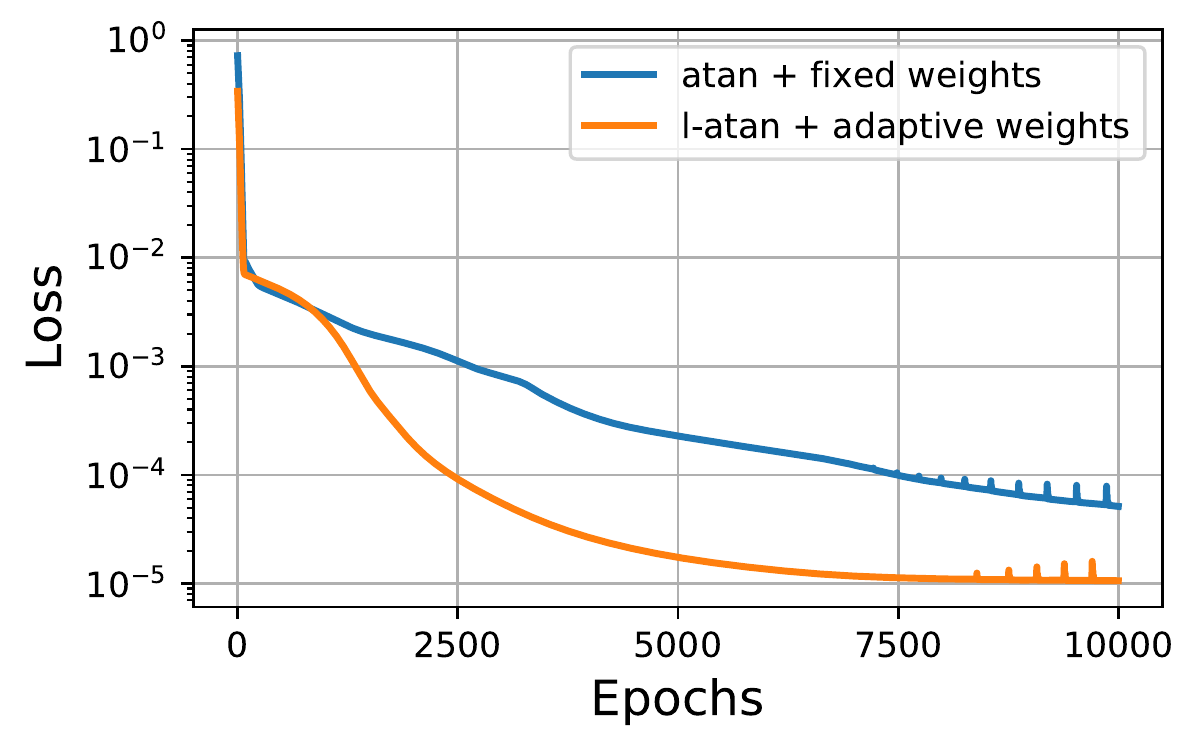}
\end{center}
\caption{
{\color{black}
A comparison of loss curves for training of the velocity model shown in Figure~\ref{fig:vofz_velmodel} using the arctangent activation function with fixed weights in the loss function (blue) compared to the training performed using the locally adaptive arctangent activation function with adaptive weights for the loss terms (orange).
}
}%
\label{fig:vofz_loss}
\end{figure}

\begin{figure}[ht!]
\begin{center}
\subfigure[]{%
\label{fig:vofz_pinnerror}
\includegraphics[width=0.43\textwidth]{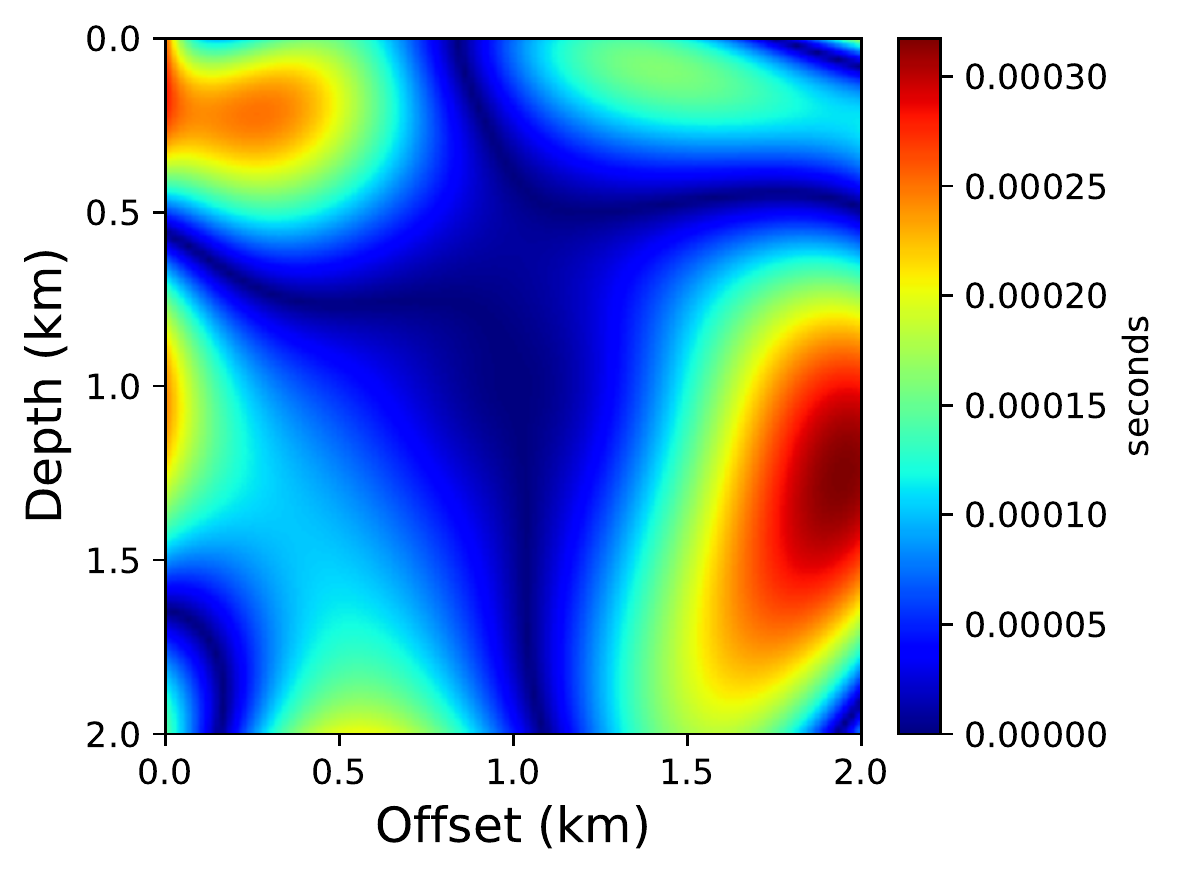}
}
\subfigure[]{%
\label{fig:vofz_fsmerror}
\includegraphics[width=0.43\textwidth]{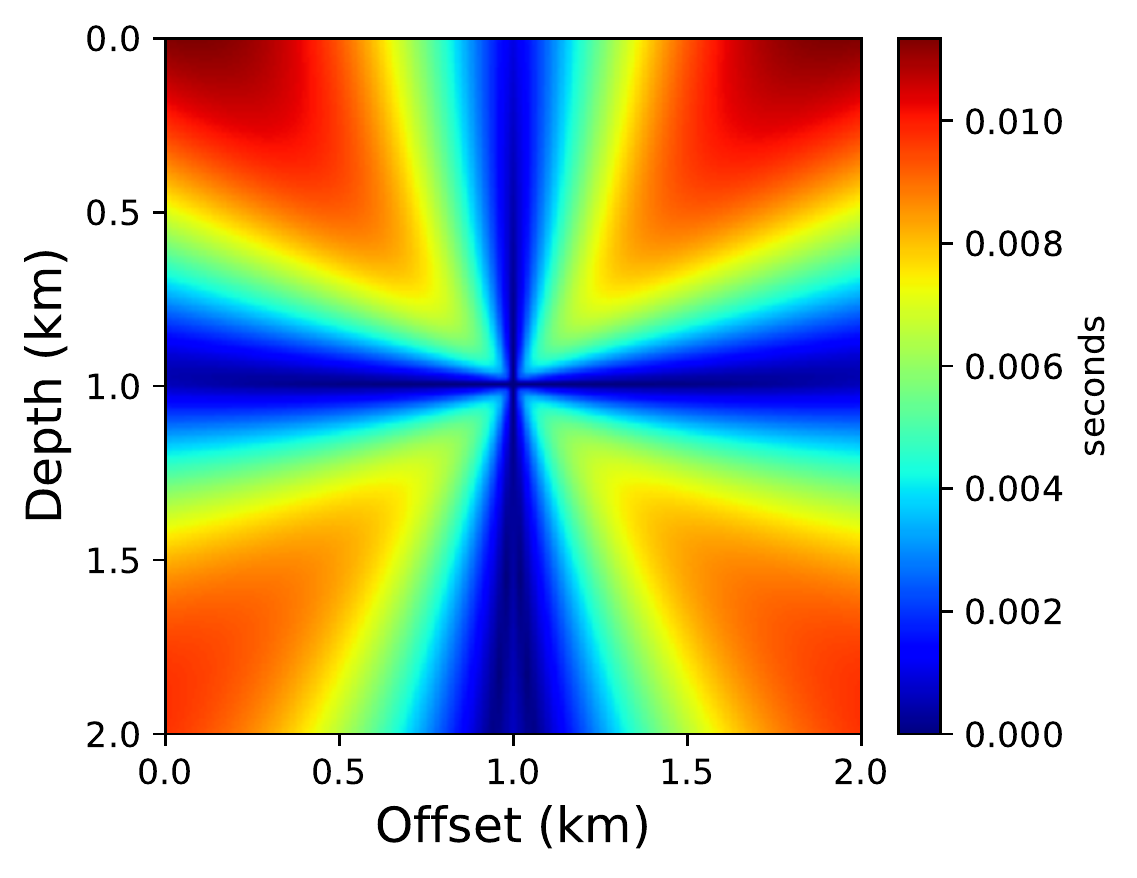}
}
\end{center}
\caption{
The absolute traveltime errors for the PINN eikonal solution (a) and the first-order fast sweeping solution (b) for the velocity model and the source location shown in Figure~\ref{fig:vofz_velmodel}. 
}%
\label{fig:vofz_errors}
\end{figure}

\begin{figure}[ht!]
\begin{center}
\includegraphics[width=0.35\textwidth]{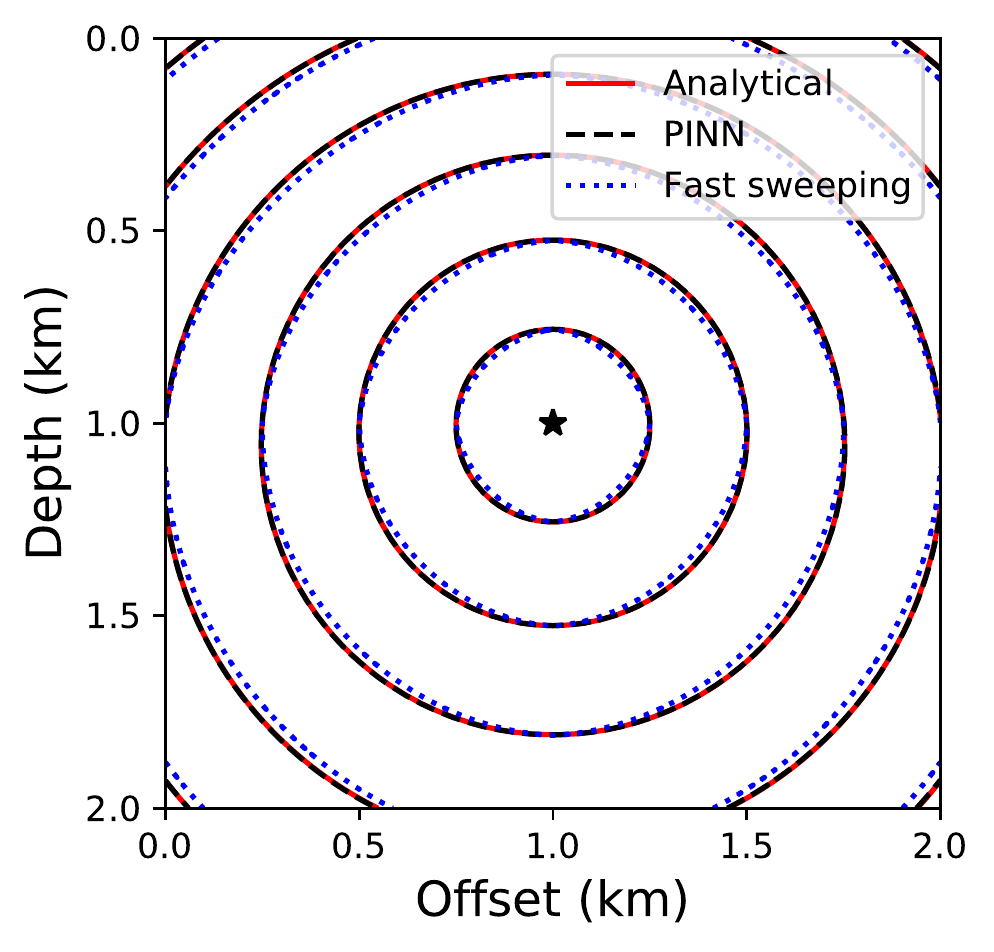}
\end{center}
\caption{
Traveltime contours for the analytical solution (red), PINN eikonal solution (dashed black), and the first-order fast sweeping solution (dotted blue). The velocity model and the source location considered are shown in Figure~\ref{fig:vofz_velmodel}.
}%
\label{fig:vofz_contours}
\end{figure}

\begin{figure}[ht!]
\begin{center}
\includegraphics[width=0.42\textwidth]{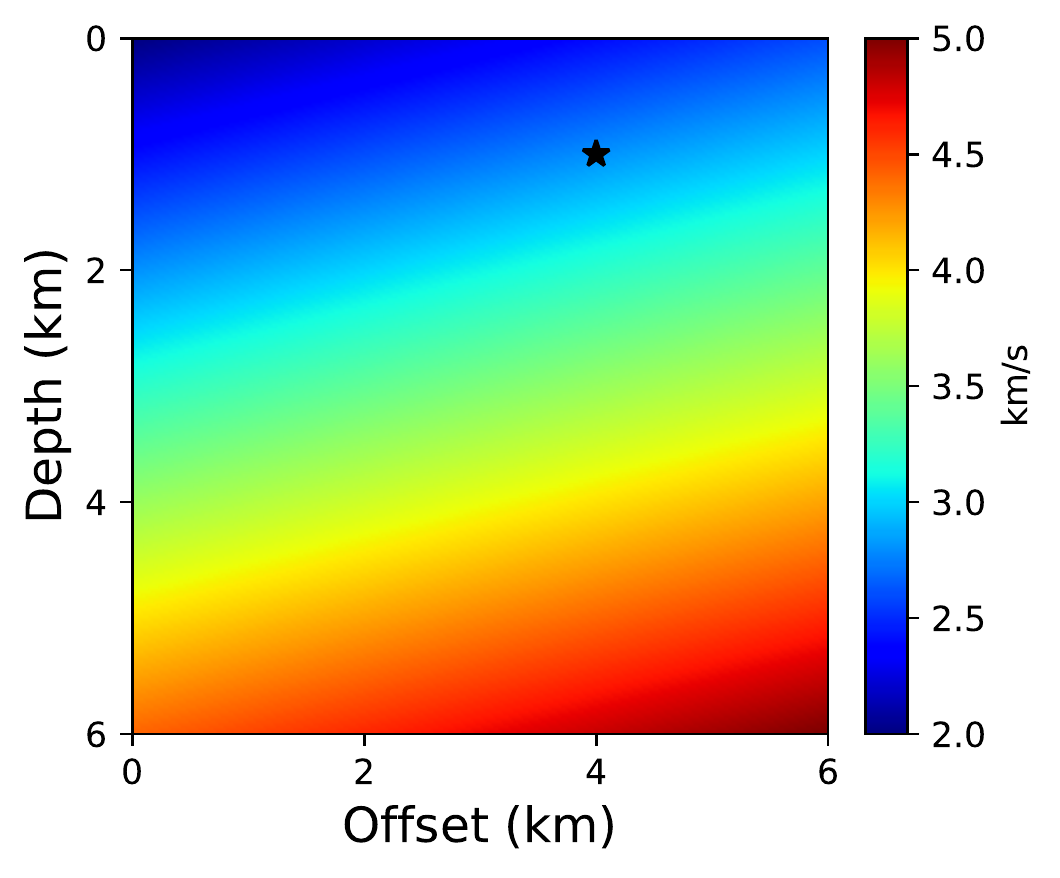}
\end{center}
\caption{
A velocity model with a constant vertical velocity gradient of 0.4~$\text{s}^{-1}$ and a horizontal velocity gradient of 0.1~$\text{s}^{-1}$. Black star indicates the point-source location used for the test.
}%
\label{fig:vofxz_velmodel}
\end{figure}

\begin{figure}[ht!]
\begin{center}
\includegraphics[width=0.46\textwidth]{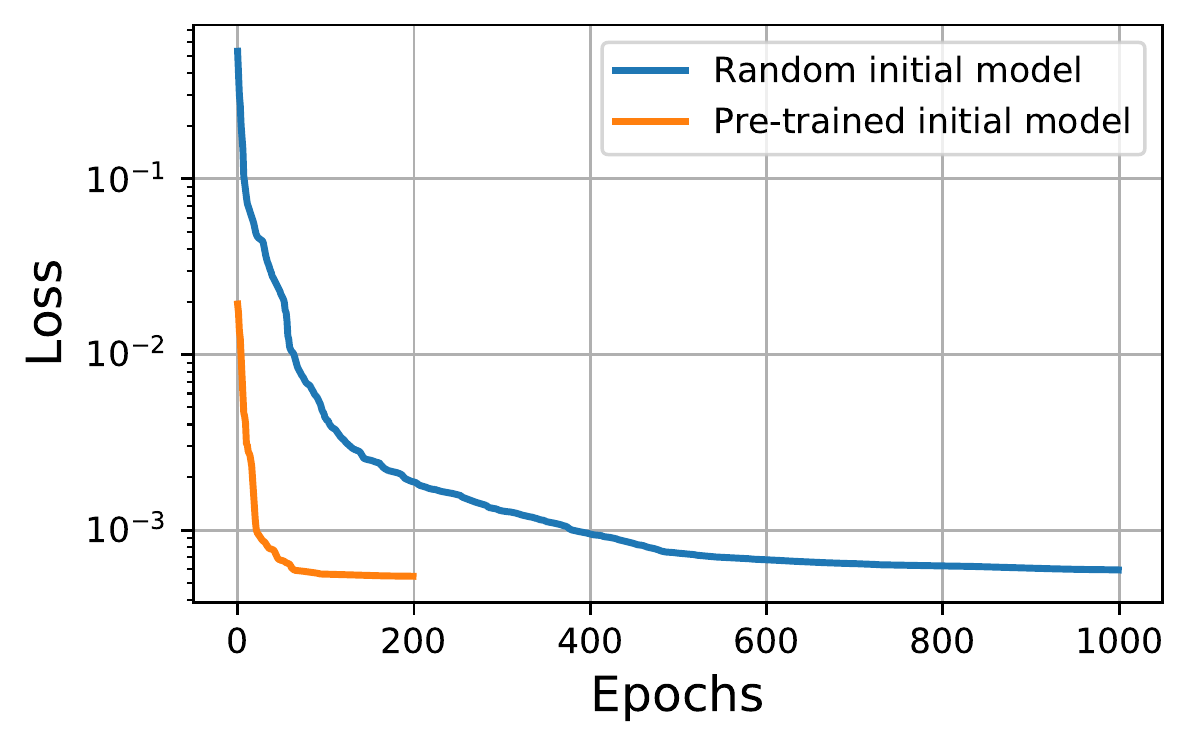}
\end{center}
\caption{
A comparison of convergence history for training of the velocity model shown in Figure~\ref{fig:vofxz_velmodel} using a randomly initialized model (blue) and a pre-trained initial model (orange) indicating the number of epochs 
needed for convergence. 
}%
\label{fig:vofxz_loss}
\end{figure}

{\color{black}
In Figure~\ref{fig:vofz_loss}, we demonstrate the efficacy of using a locally adaptive activation function~\cite{jagtap2020locally} and the adaptive weighting strategy of the terms in the loss function \cite{wang2020understanding} compared to the standard PINN approach. For both cases, the network is trained on 25\% of the total grid points, randomly selected, using the Adam optimizer for 10,000 epochs. We observe markedly improved convergence by using the locally adaptive arctangent activation function and adaptive weights for the loss terms. Therefore, all examples reported in this study utilize these techniques to accelerate the convergence of the PINN model.}

In Figure~\ref{fig:vofz_pinnerror}, we show the absolute traveltime errors for the PINN eikonal solver considering the same velocity model and source position. 
Once the network is trained, we evaluate the network on the same 101~$\times$~101 regular grid. For comparison, we also plot absolute traveltime errors for the first-order fast sweeping solution in Figure~\ref{fig:vofz_fsmerror} on the same grid. We observe that despite using only 25\% of the total grid points for training, the PINN eikonal solution is significantly more accurate than the first-order fast sweeping solution. As can be seen in Figure~\ref{fig:vofz_fsmerror}, the fast sweeping solution suffers from large errors in the diagonal direction, whereas the errors for the PINN eikonal solver are more randomly distributed. We also plot traveltime contours in Figure~\ref{fig:vofz_contours} comparing the analytical solution with the PINN eikonal solution and the first-order fast sweeping solution visually.

Next, we investigate the applicability of transfer learning to the PINN eikonal solver. Transfer learning is a machine learning technique that relies on storing knowledge gained while solving one problem and applying it to a different but related problem. We explore if the network trained on the previous example can be used to compute the solution for a different source location and velocity model. To this end, we consider a 6$\times$6~km$^2$ velocity model with a vertical gradient of 0.4~$\text{s}^{-1}$ and a horizontal gradient of 0.1~$\text{s}^{-1}$. The point-source is also relocated to $(4~\text{km},1~\text{km})$ as shown in Figure~\ref{fig:vofxz_velmodel}. The model is discretized on a 301~$\times$~301 grid with a grid spacing of 20~m along both axes.

To train the network for this case, instead of initializing the network with random weights, we use the weights from the network trained for the previous example. We re-train this neural network using 25\% of the total grid points, selected randomly, using the L-BFGS-B solver~\cite{zhu1997algorithm}. Starting with pre-trained weights allow us to use a super-linear optimization method for faster convergence as opposed to starting with the first-order Adam optimizer for stable convergence and then switching to the L-BFGS-B optimizer, as suggested in previous studies~\cite{raissi2019physics}. For comparison, we also train a neural network with random initialization and the loss curves are compared in Figure~\ref{fig:vofxz_loss}. Despite using a different velocity model with a relocated point-source and a larger model, we observe that the network with pre-trained weights converges much faster than the one trained from scratch. This is a highly desirable property of the PINN eikonal solver since seismic applications, such as earthquake source localization or seismic imaging, require repeated traveltime computations for multiple source locations and updated velocity models. In comparison, conventional numerical algorithms, such as the fast sweeping method, require the same computational effort for even a slight variation in the velocity model or source location, which is a major source of the computational bottleneck, particularly in large 3D models.

\begin{figure}[ht!]
\begin{center}
\subfigure[]{%
\label{fig:vofxz_pinnerror}
\includegraphics[width=0.4\textwidth]{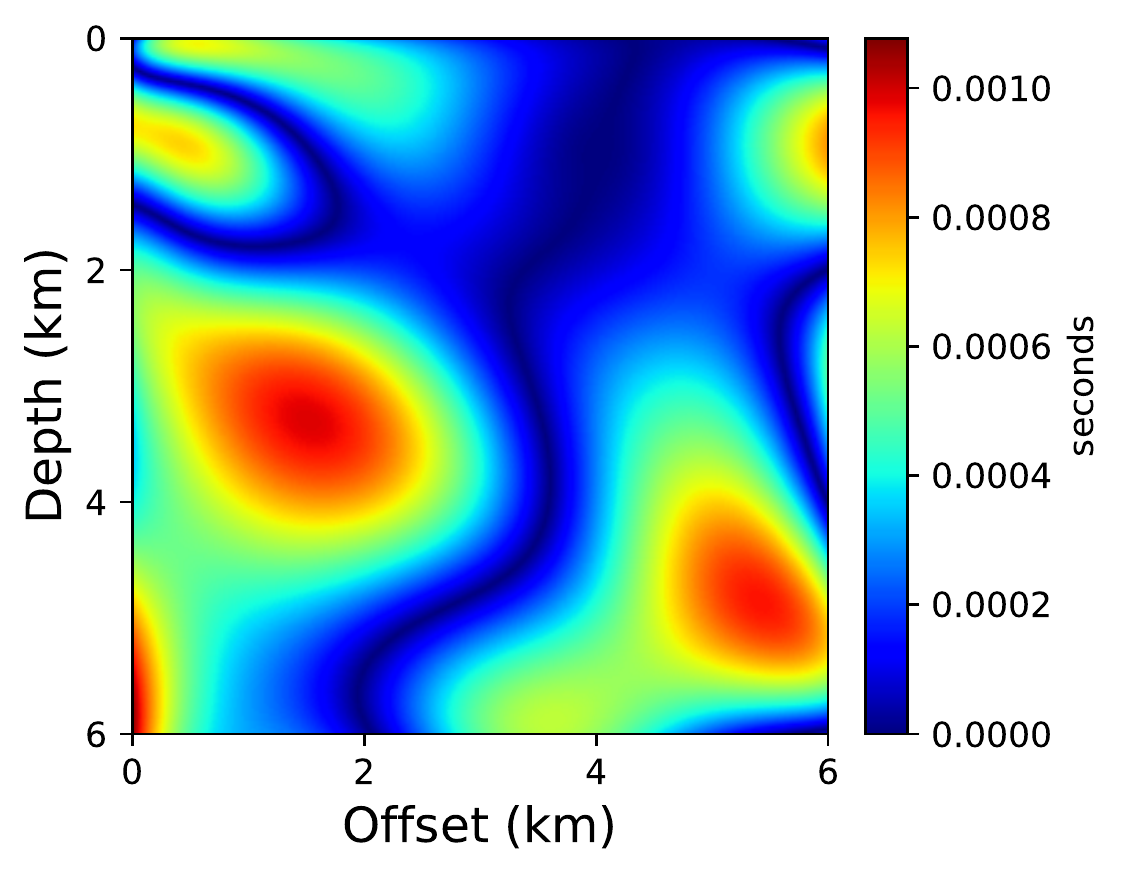}
}
\subfigure[]{%
\label{fig:vofxz_fsmerror}
\includegraphics[width=0.4\textwidth]{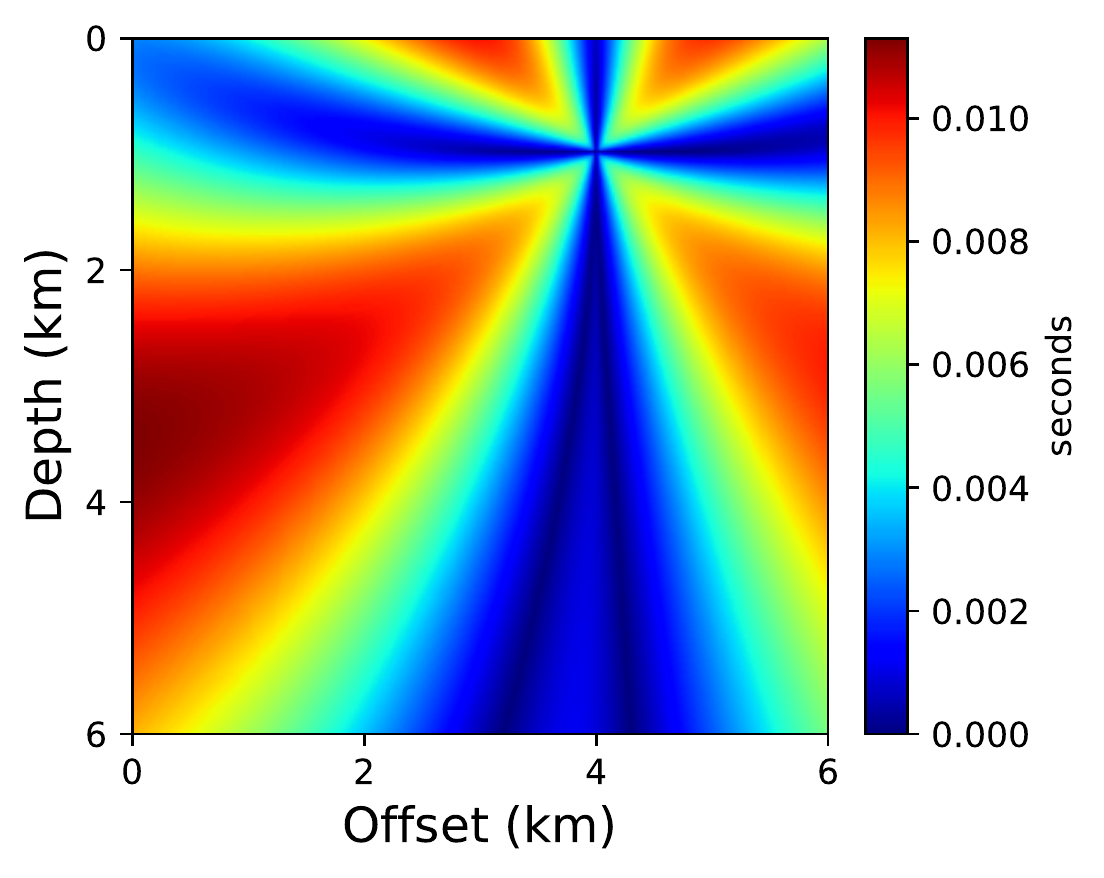}
}
\end{center}
\caption{
The absolute traveltime errors for the PINN eikonal solution (a) and the first-order fast sweeping solution (b) for the velocity model and the source location shown in Figure~\ref{fig:vofxz_velmodel}. 
}%
\label{fig:vofxz_errors}
\end{figure}

\begin{figure}[ht!]
\begin{center}
\includegraphics[width=0.35\textwidth]{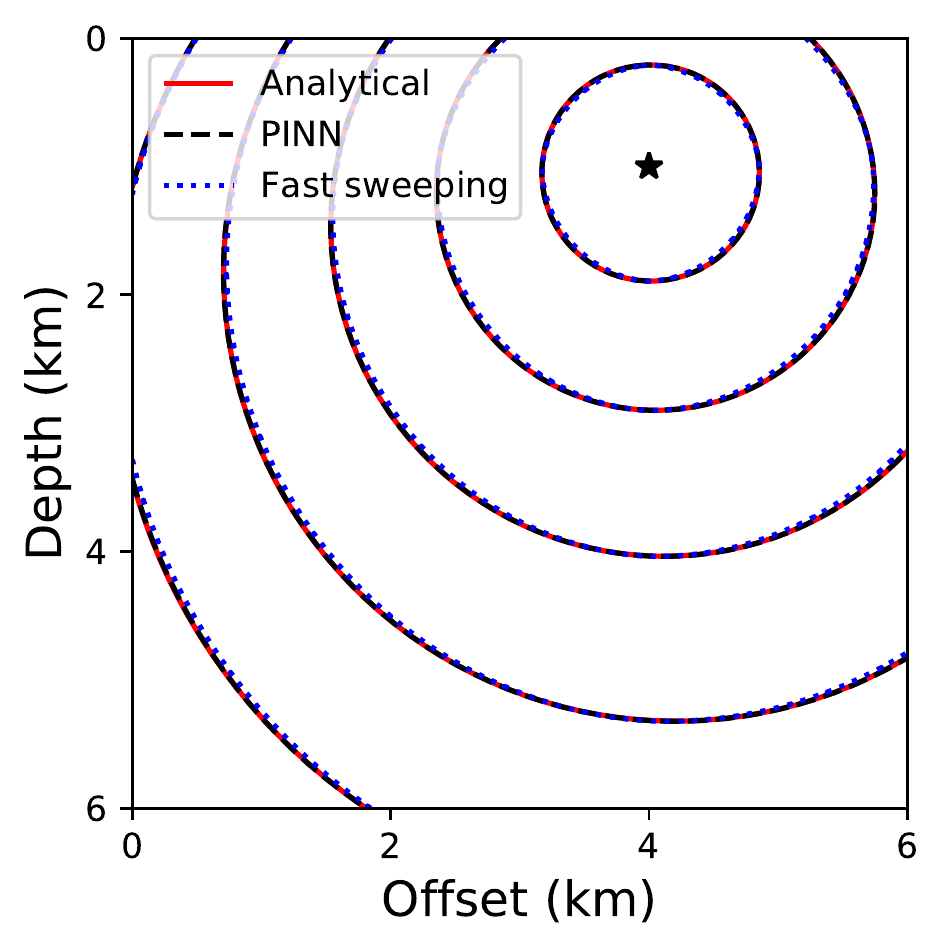}
\end{center}
\caption{
Traveltime contours for the analytical solution (red), PINN eikonal solution (dashed black), and the first-order fast sweeping solution (dotted blue). The velocity model and the source location considered are shown in Figure~\ref{fig:vofxz_velmodel}.
}%
\label{fig:vofxz_contours}
\end{figure}

\begin{figure}[ht!]
\begin{center}
\includegraphics[width=0.42\textwidth]{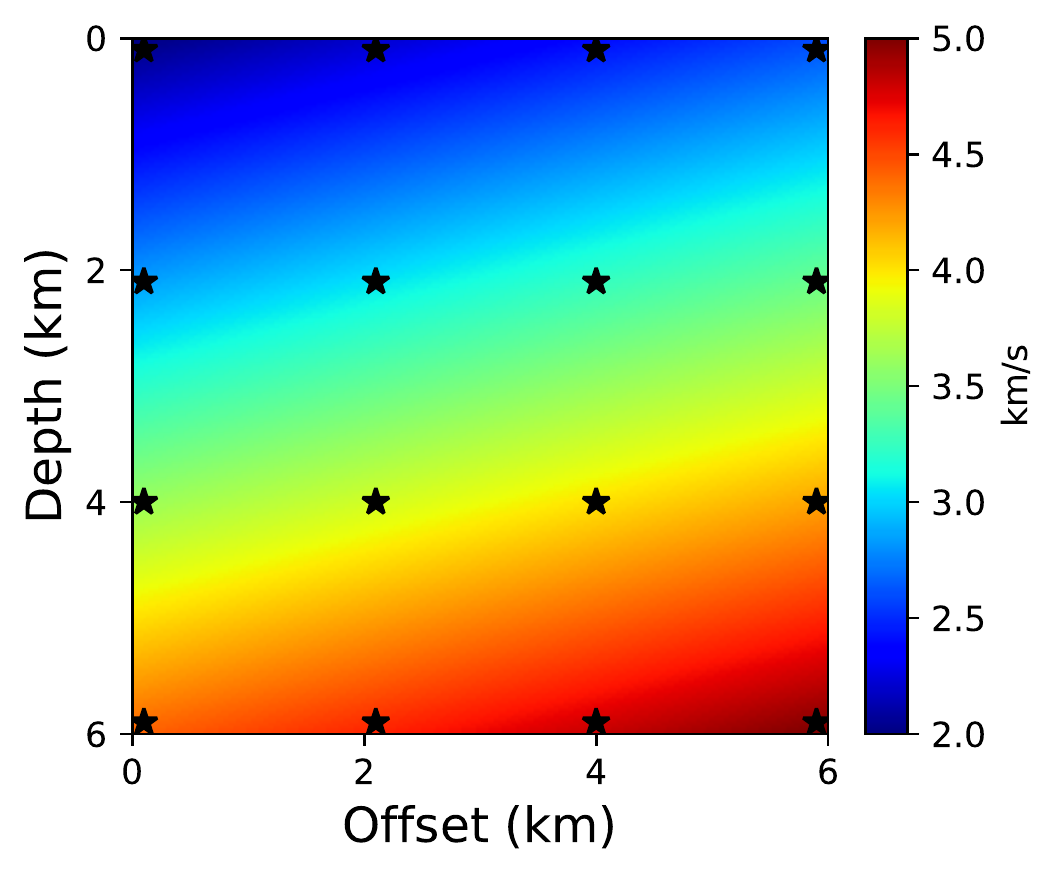}
\end{center}
\caption{
A velocity model with a constant vertical velocity gradient of 0.4~$\text{s}^{-1}$ and a horizontal velocity gradient of 0.1~$\text{s}^{-1}$. Black stars indicate locations of sources used to train the network as a surrogate model.
}%
\label{fig:vofxz_surrogate_sources}
\end{figure}

Figure~\ref{fig:vofxz_errors} compares the absolute traveltime errors computed using the PINN eikonal solution with the pre-trained initial model and the first-order fast sweeping solution. We observe that despite using significantly fewer epochs, the solution accuracy is not compromised. The traveltime contours, shown in Figure~\ref{fig:vofxz_contours}, confirm this observation visually.

{\color{black} Next, we explore if a PINN model trained on solutions computed for various source locations in a given velocity model can be used as a surrogate model. To do this, the neural network is modified to include the source location $\mathbf{x_s} = (x_s,z_s)$ as inputs in addition to the grid points $\mathbf{x} = (x,z)$. We train the network on solutions computed for 16 sources located at regular intervals for the same model, as shown in Figure~\ref{fig:vofxz_surrogate_sources}. These computed solutions act as additional data points for training the surrogate model and these points along with the computed solution can be lumped with the boundary term in the loss function. Through this training process, the network learns the mapping between a considered source location $\mathbf{x_s}$ and a point in the model space $\mathbf{x}$ to the corresponding traveltime factor value $\hat{\tau}(\mathbf{x_s},\mathbf{x})$. 
Once the surrogate model is trained with source locations as additional input parameters, the traveltime function for new source locations for the given velocity model can be computed rapidly with just a single evaluation of the network. This is similar to obtaining an analytic solver as no further training is needed for computing traveltimes corresponding to additional source locations. This property is particularly advantageous for large 3D models that need thousands of such computations for inversion applications.}

\begin{figure}[ht!]
\begin{center}
\subfigure[]{%
\label{fig:vofxz_surrogate_pinnerror}
\includegraphics[width=0.4\textwidth]{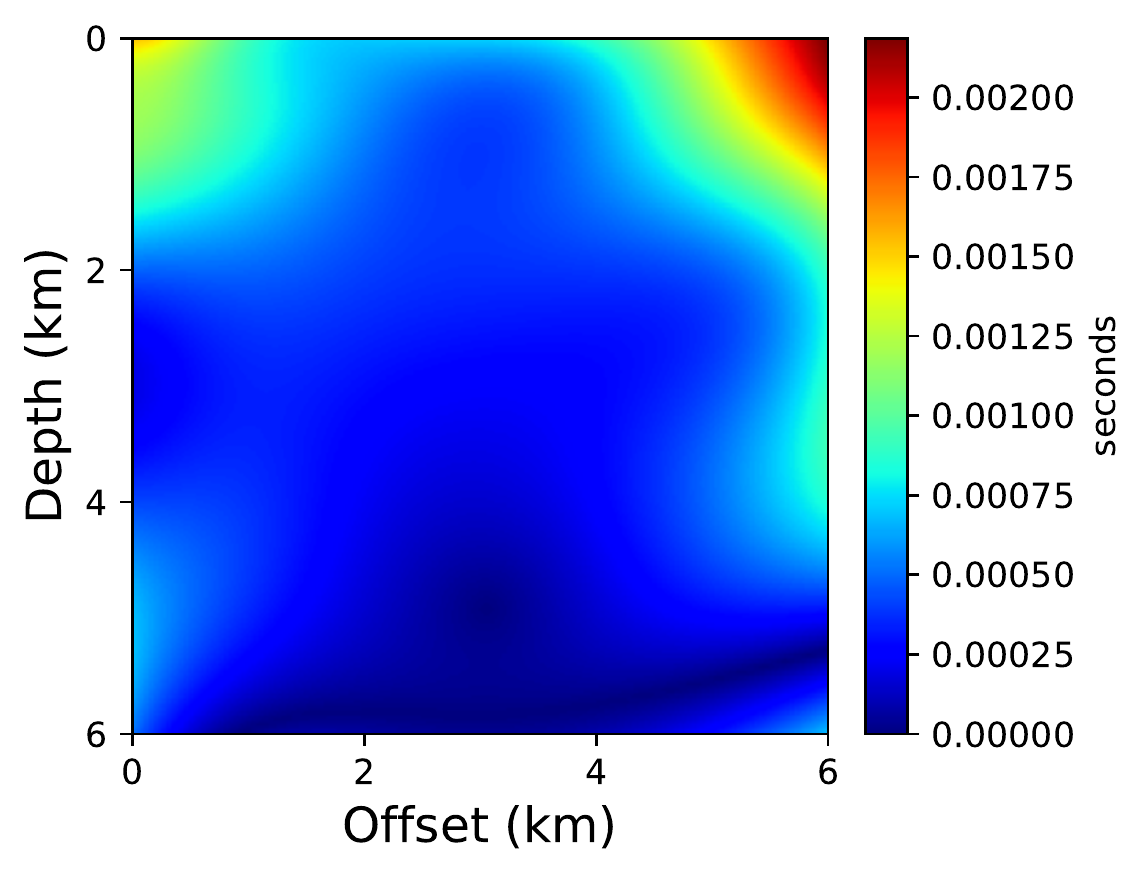}
}
\subfigure[]{%
\label{fig:vofxz_surrogate_fsmerror}
\includegraphics[width=0.4\textwidth]{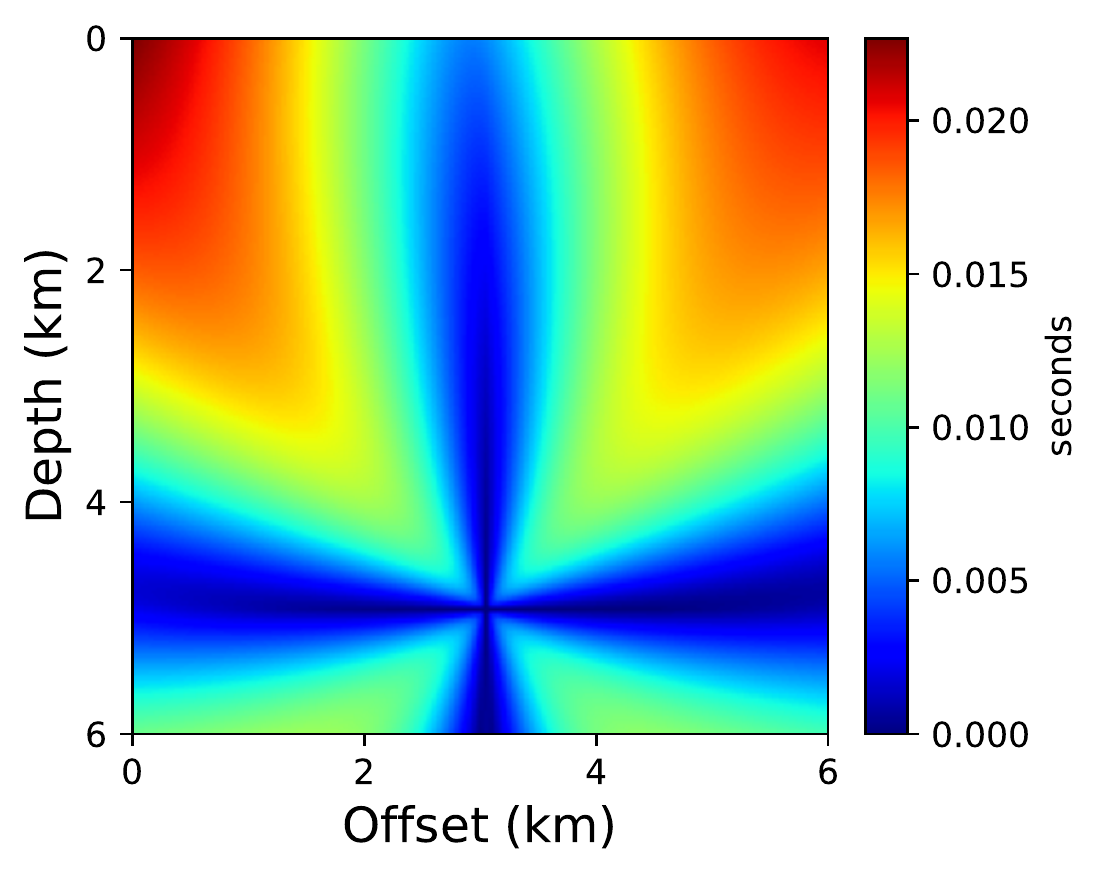}
}
\end{center}
\caption{
{\color{black}
The absolute traveltime errors for the solution computed using the PINN surrogate model (a) and the first-order fast sweeping solution (b) for the velocity model shown in Figure~\ref{fig:vofxz_surrogate_sources} and the source is located at (3.05~km, 4.95~km).}
}%
\label{fig:vofxz_surrogate_errors}
\end{figure}

\begin{figure}[ht!]
\begin{center}
\includegraphics[width=0.33\textwidth]{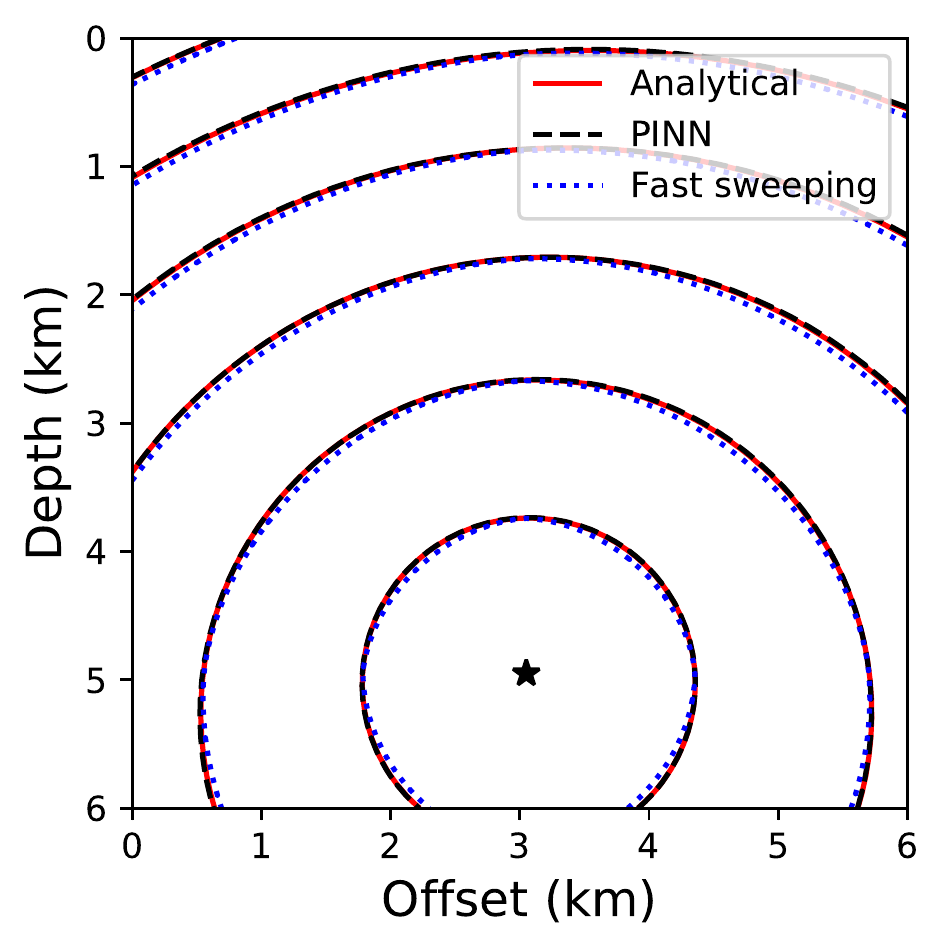}
\end{center}
\caption{
{\color{black}
Traveltime contours for solutions obtained using the analytical formula (red), the PINN surrogate model (dashed black), and the first-order fast sweeping solver (dotted blue). The velocity model considered is shown in Figure~\ref{fig:vofxz_surrogate_sources} and a source located at (3.05~km, 4.95~km).} 
}%
\label{fig:vofxz_surrogate_contours}
\end{figure}

\begin{figure}[ht!]
\begin{center}
\subfigure[]{%
\label{fig:vtiseam_vz}
\includegraphics[width=0.4\textwidth]{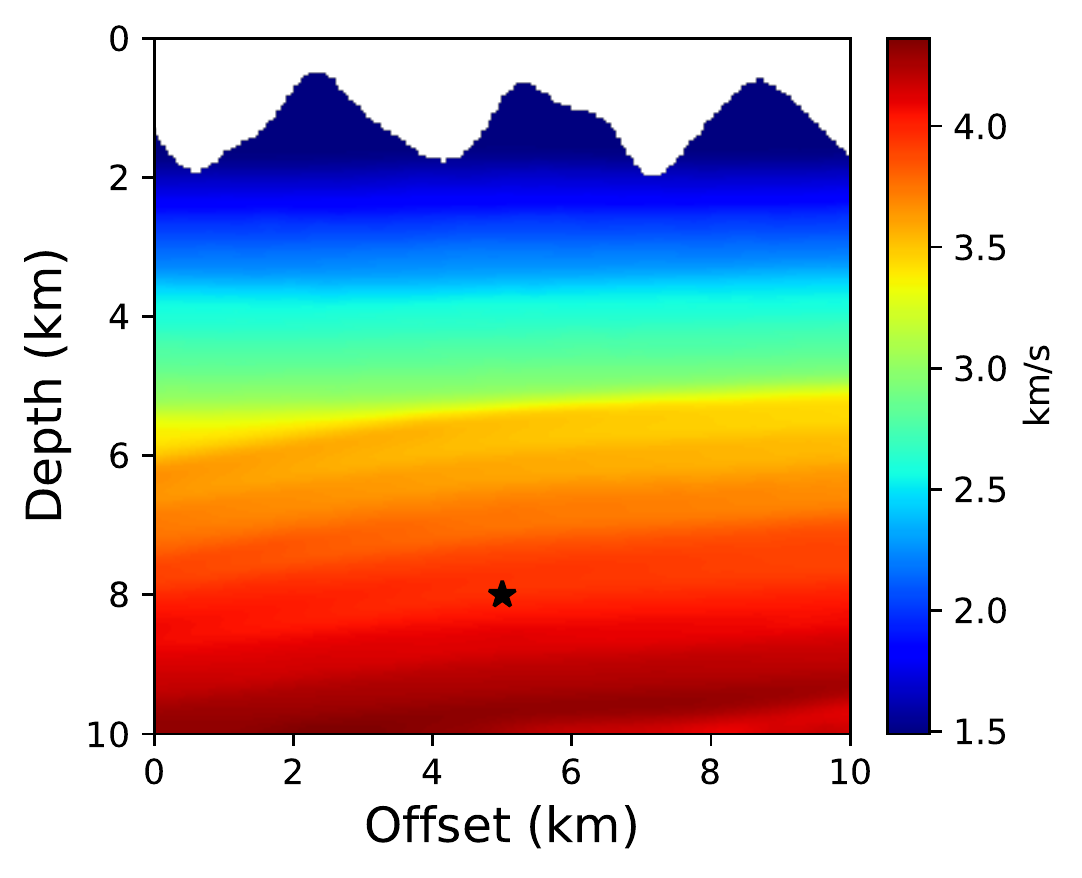}
}
\subfigure[]{%
\label{fig:vtiseam_vx}
\includegraphics[width=0.4\textwidth]{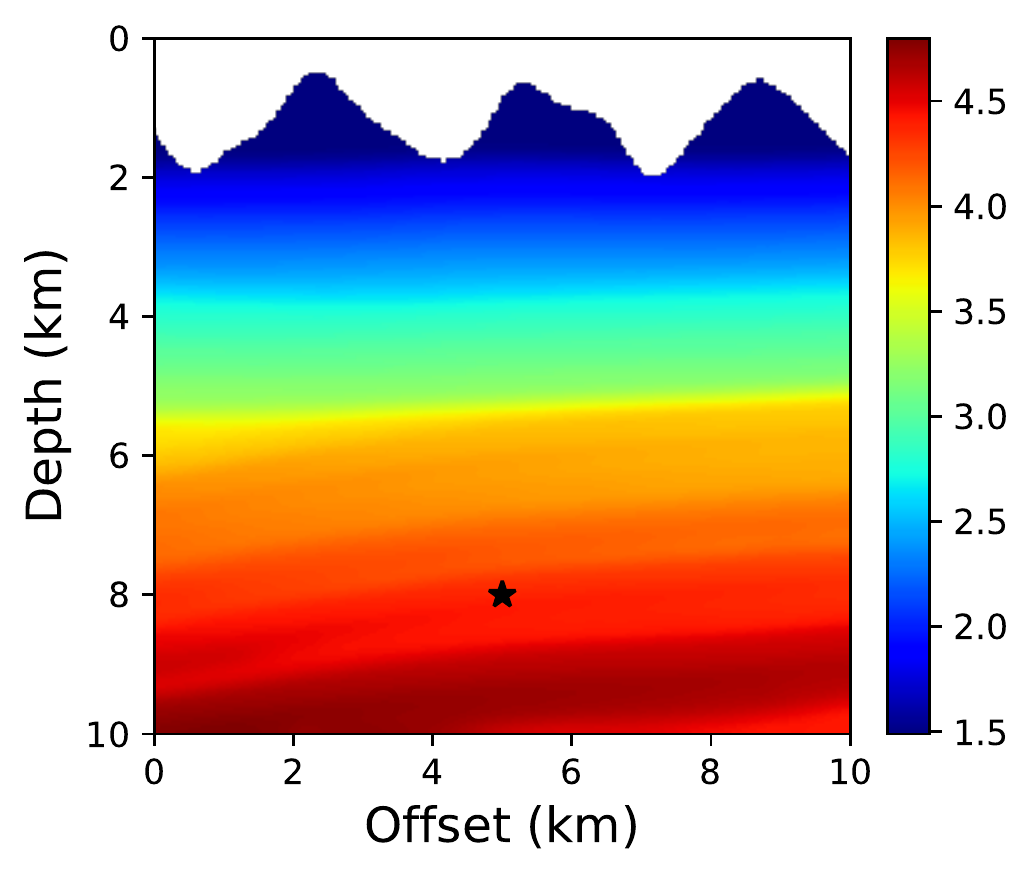}
}
\end{center}
\caption{
The vertical (a) and horizontal (b) velocity models for the elliptically anisotropy model with irregular topography. Black star indicates the position of the point-source. 
}%
\label{fig:vtiseam_model}
\end{figure}

To demonstrate if the surrogate model approach yields an accurate solution, we use the trained surrogate model to compute traveltime solution for a {\color{black} source located at (3.05~km, 4.95~km). This source position is deliberately chosen to be the furthest away from the training source locations to better analyze the accuracy limits of the surrogate model.} We can confirm by looking at the absolute traveltime errors, shown in Figure~\ref{fig:vofxz_surrogate_errors}, that the trained surrogate model yields a highly accurate solution compared to the first-order fast sweeping solution even though no additional training is performed for this randomly chosen source point (see Figure~\ref{fig:vofxz_surrogate_contours} for traveltime contours).

Moreover, transfer learning can be used to efficiently build surrogate models for updated velocity models, i.e., by initializing the PINN surrogate model for the updated velocities using weights from the already trained surrogate model. Therefore, the transfer learning technique combined with surrogate modeling can be used to build a highly efficient traveltime modeling engine for seismic inversion compared to conventional algorithms that do not afford such flexibility.

\begin{figure}[ht!]
\begin{center}
\subfigure[]{%
\label{fig:vtiseam_pinerror}
\includegraphics[width=0.4\textwidth]{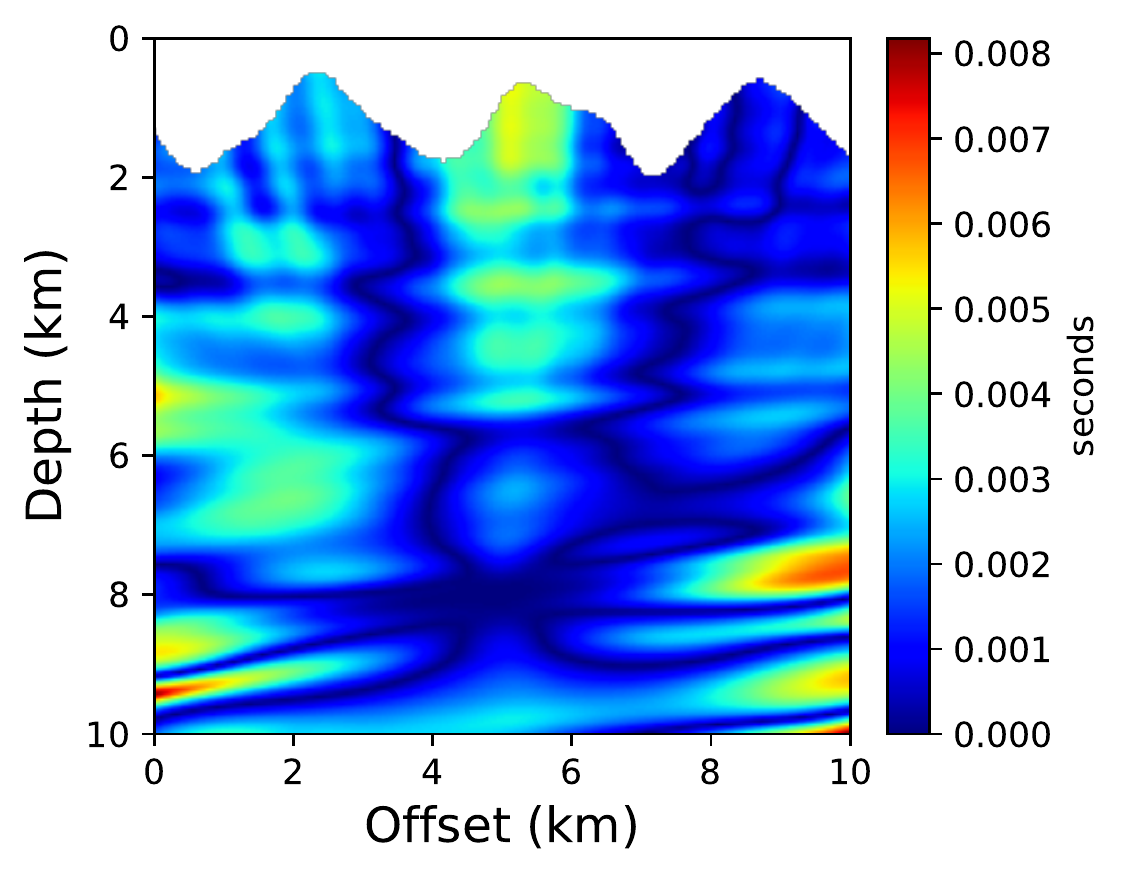}
}
\subfigure[]{%
\label{fig:vtiseam_fsmerror}
\includegraphics[width=0.4\textwidth]{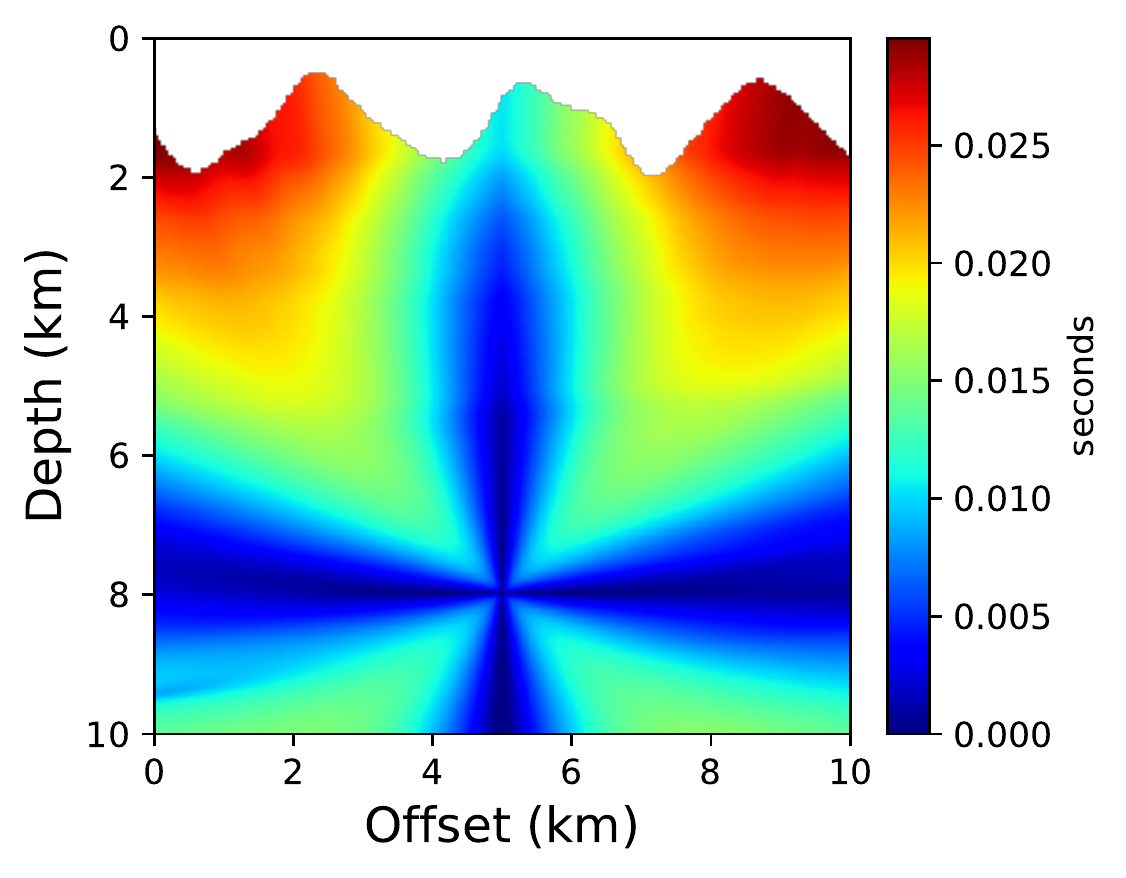}
}
\end{center}
\caption{
The absolute traveltime errors for the PINN eikonal (a) and first-order fast sweeping (b) solutions for the anisotropic model and the source location shown in Figure~\ref{fig:vtiseam_model}. 
}%
\label{fig:vtiseam_error}
\end{figure}

\begin{figure}[ht!]
\begin{center}
\includegraphics[width=0.35\textwidth]{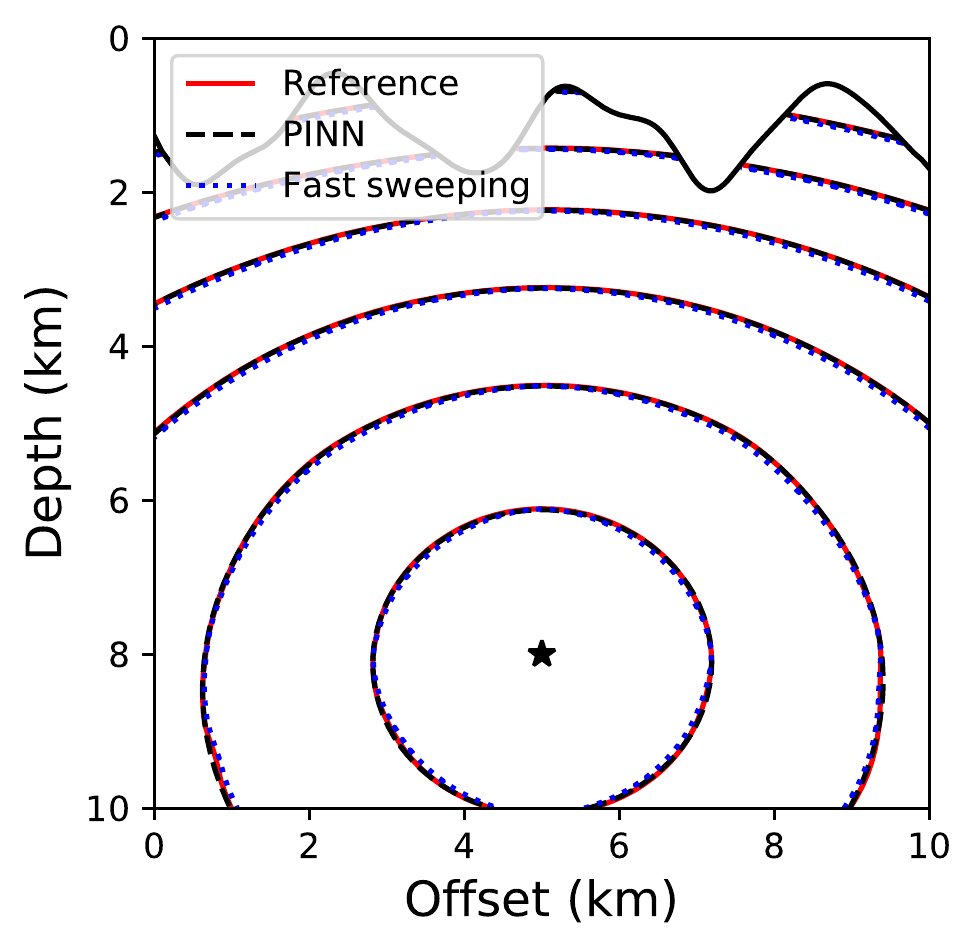}
\end{center}
\caption{
Traveltime contours for the reference solution (red), PINN eikonal solution (dashed black), and the first-order fast sweeping solution (dotted blue). The velocity model and the source location considered are shown in Figure~\ref{fig:vtiseam_model}. The black solid curve indicates the surface topography.
}%
\label{fig:vtiseam_contours}
\end{figure}

\begin{figure}[ht!]
\begin{center}
\includegraphics[width=0.44\textwidth]{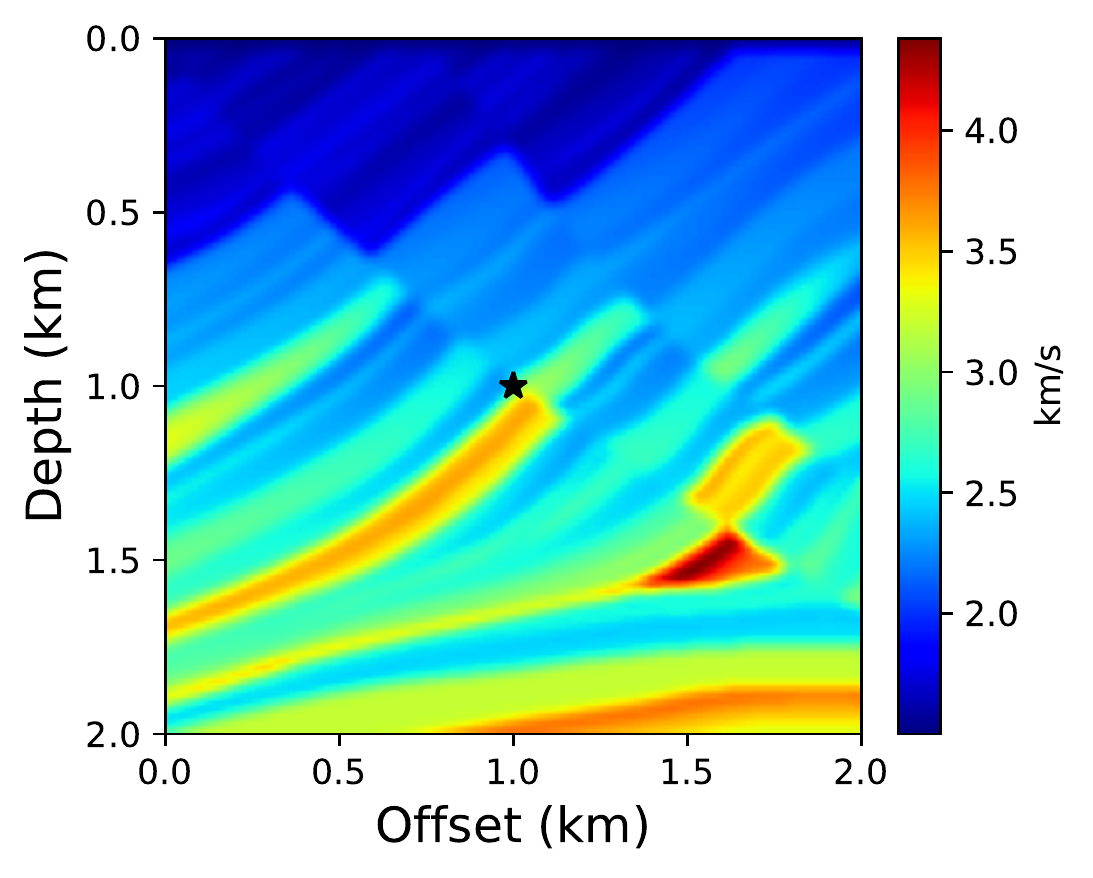}
\end{center}
\caption{
A highly heterogeneous portion of the Marmousi model used for traveltime computation. Black star indicates the point-source location used for the test.
}%
\label{fig:marm_velmodel}
\end{figure}

Next, to demonstrate the flexibility of the proposed framework, we consider an anisotropic model with irregular topography. We consider the simplest form of anisotropy, known as elliptical anisotropy, which uses a vertical and a horizontal velocity to parameterize the model and to define an elliptical phase velocity surface. The considered model parameters are shown in Figure~\ref{fig:vtiseam_model} with a rugged topography layer on top.
The introduction of anisotropy necessitates the construction of a new solver for the fast sweeping method, whereas for PINNs, it requires only an update to the loss function term that embeds the residual of the PDE. Moreover, for conventional methods, the presence of non-uniform topography requires special treatment, such as mathematical flattening of the free-surface and solving a topography-dependent eikonal equation~\cite{lan2013topography}. This not only adds to the complexity of the eikonal solver but also results in a considerable increase in the computational cost. On the contrary, being a mesh-free method, PINNs do not require any special treatment. For models with irregular topography, only the grid points below the free-surface are used for training and evaluation of the network.

\begin{figure}[ht!]
\begin{center}
\subfigure[]{%
\label{fig:marm_pinnerror}
\includegraphics[width=0.43\textwidth]{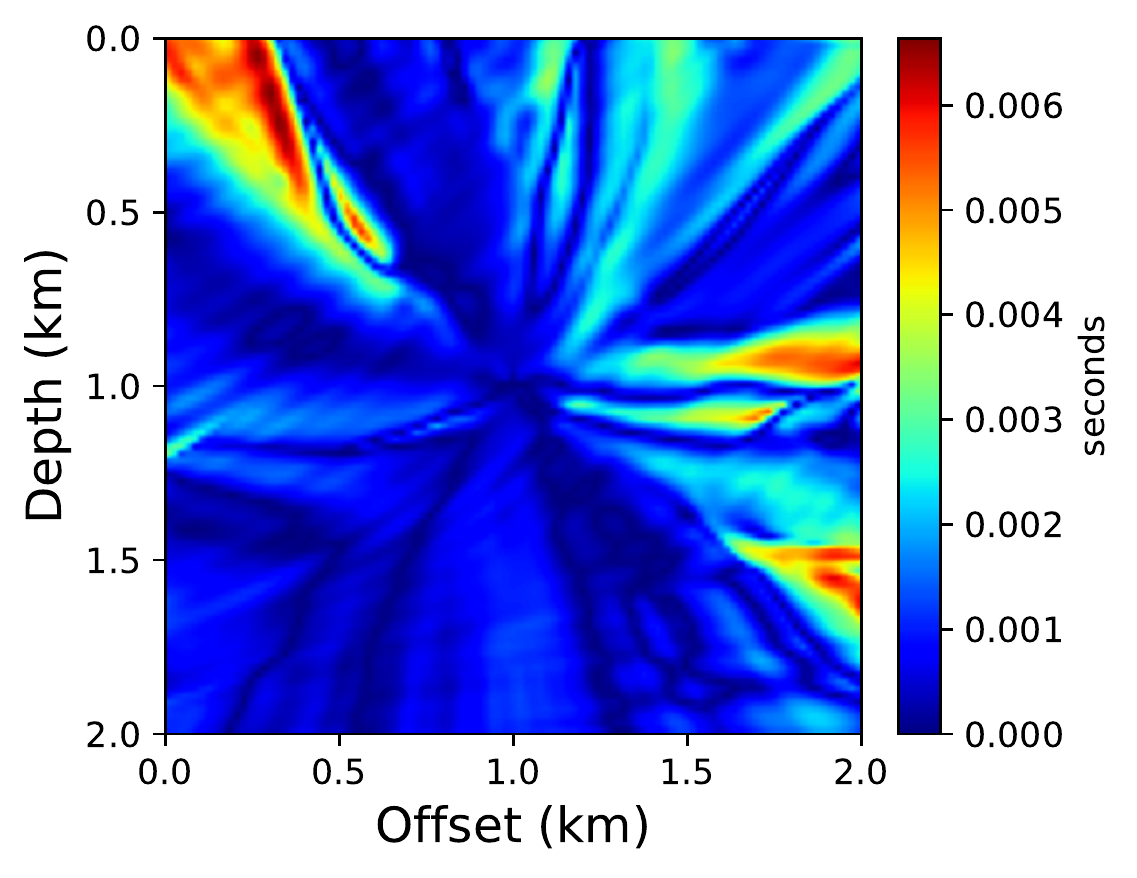}
}
\subfigure[]{%
\label{fig:marm_fsmerror}
\includegraphics[width=0.43\textwidth]{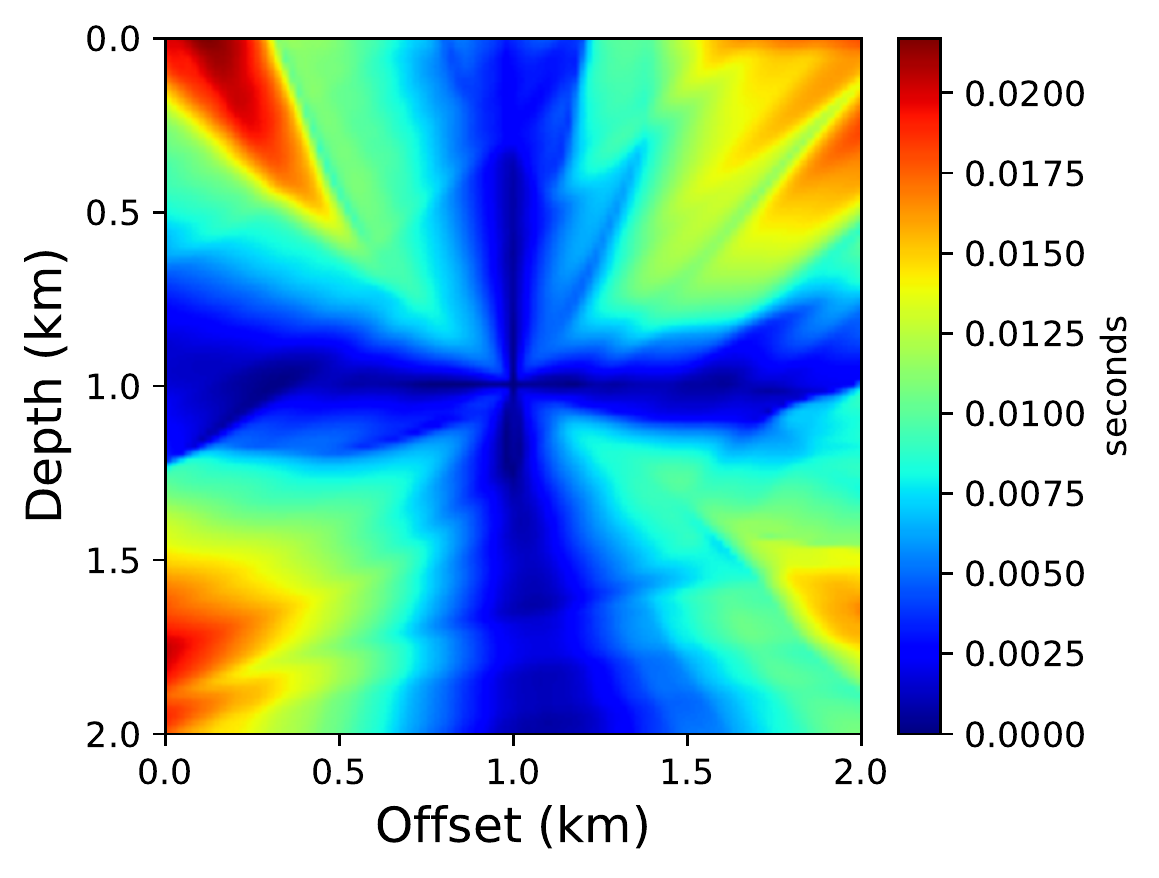}
}
\end{center}
\caption{
The absolute traveltime errors for the PINN eikonal solution (a) and the first-order fast sweeping solution (b) for the velocity model and the source location shown in Figure~\ref{fig:marm_velmodel}. 
}%
\label{fig:marm_errors}
\end{figure}

\begin{figure}[ht!]
\begin{center}
\includegraphics[width=0.35\textwidth]{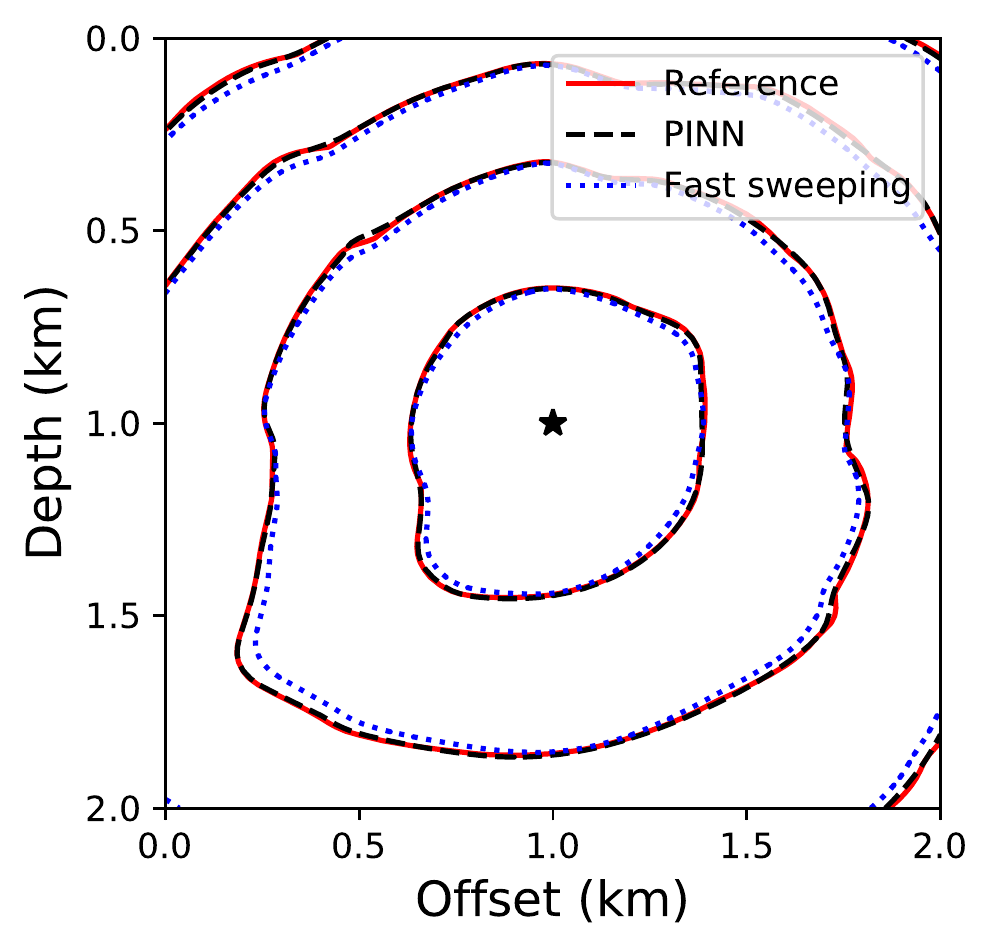}
\end{center}
\caption{
Traveltime contours for the reference solution (red), PINN eikonal solution (dashed black), and the first-order fast sweeping solution (dotted blue). The velocity model and the source location considered are shown in Figure~\ref{fig:marm_velmodel}. 
}%
\label{fig:marm_contours}
\end{figure}

Finally, we test the PINN eikonal solver on a highly heterogeneous portion of the Marmousi model as shown in Figure~\ref{fig:marm_velmodel}. We consider a source located at $(1~\text{km},1~\text{km})$. This is a particularly challenging model due to sharp velocity variations. The model is discretized on a 101~$\times$~101 grid with a spacing of 20~m along both axes. Starting with the pre-trained weights from the model shown in Figure~\ref{fig:vofz_velmodel}, we train the network on $30\%$ of the total grids point, randomly selected from the discretized computational domain. The training is done using an L-BFGS-B solver for 12,000 epochs. Once the network is trained, we evaluate it on the same 101~$\times$~101 regular grid. For comparison, we also compute the traveltime solution using the first-order fast sweeping method on the same regular grid. The absolute traveltime errors for both the approaches are compared in Figure~\ref{fig:marm_errors} and the traveltime contours are shown in Figure~\ref{fig:marm_contours}. We observe significantly better accuracy for the PINN eikonal solver compared to the first-order fast sweeping method. 

Despite the model being anisotropic in nature, we initialize the PINN model using pre-trained weights from the isotropic model shown in Figure~\ref{fig:vofz_velmodel}. The velocity model is discretized using a grid spacing of 50~m along both axes. The training is done using only 12\% of the total grid points, selected randomly, below the free-surface using L-BFGS-B optimizer for 2000 epochs. For a source located at (5~km, 8~km), the absolute traveltime errors for the PINN and fast sweeping solvers are shown in Figure~\ref{fig:vtiseam_error}. The corresponding traveltime contours are shown in Figure~\ref{fig:vtiseam_contours}. Again, we observe better accuracy for PINNs compared to the first-order fast sweeping method.  

However, compared to previous examples, we also observe considerably slower convergence for the Marmousi model. This can be attributed to the spectral bias of fully-connected neural networks~\cite{rahaman2019spectral}, which underscores the learning bias of deep neural networks towards smoother representations. Since the solution for the Marmousi model contains plenty of local fluctuations or high-frequency features compared to prior examples, this means a longer training time for the neural network to accurately capture the underlying local features. While the use of locally adaptive activation functions, gradient balancing, transfer learning, and a second-order optimizer help improve the convergence rate, further advances are needed to make PINNs computationally feasible for such highly heterogeneous velocity models.


\section{Discussion}

In a conventional deep learning application, a neural network is trained by minimizing a loss function that typically measures the mismatch between the network's predicted outputs and their expected (true) values, also known as training data. However, there are several limitations associated with such models that solely rely on a labeled dataset and are oblivious to the scientific principles governing real-world phenomena. For cases when the available training and test data are insufficient, such models often learn spurious relationships that are misleading. However, the biggest concern of such a data-driven model is the lack of scientific consistency of their predictions to known physical laws that have been the cornerstone of knowledge discovery across scientific disciplines for centuries. 

A case in point is the failure of Google Flu Trends~--~a system designed to predict the onset of flu solely based on Google search queries without taking into account the physical knowledge of the disease spread. Despite its success in the initial years that were used to train the model, it soon started overestimating by several factors to the point that it was eventually taken down~\cite{lazer2014parable}. Such problems with black-box data science methods on scientific problems have been reported in several other publications~\cite{caldwell2014statistical,marcus2014eight,karpatne2017theory}. Furthermore, consider a neural network with \emph{rectified linear unit (ReLU)} activation function. These networks show excellent training and convergence characteristics for data-driven setups. However, it is trivial that the first spatial or temporal derivative of this network is discontinuous while the second derivative is identically zero. Considering that most physical phenomena are derived by gradients, trivially such networks cannot show generalization capabilities. In a physics-informed setup, these networks cannot be trained at all because the governing PDEs are not satisfied. 

Therefore, in the present work, we propose an eikonal solver based on the framework of physics-informed neural networks~\cite{raissi2019physics}. We leverage the capabilities of neural networks as universal function approximators~\cite{hornik1989multilayer} and define a loss function to minimize the residual of the governing eikonal equation at a chosen set of training points. This is achieved with a simple feed-forward neural network leveraging the concept of automatic differentiation~\cite{baydin2017automatic}. Through numerical tests, we observe that the proposed algorithm yield sufficiently accurate traveltimes for most seismic applications of interest. We demonstrate this by comparing the accuracy of the proposed approach against the first-order fast sweeping solution, which is a popular numerical algorithm for solving the eikonal equation.


We observe that the transfer learning technique can be used to speed up the convergence of the network for new source locations and/or updated velocity models by initializing the PINN model with the weights of a previously trained network. Moreover, having computed solutions corresponding to a few sources for a given velocity model, we can also build a surrogate model with respect to the source locations by adding them as input parameters. This essentially means that this surrogate model can then be used to compute traveltime fields corresponding to new source locations for the same velocity model just by a single evaluation of the network. These observations effectively demonstrate the potential of the proposed approach in massively speeding up many seismic applications that rely on repeated traveltime computations for multiple source locations and velocity models. 

For a rudimentary analysis of the computational cost, we note that the training of the surrogate model, shown in Figure~\ref{fig:vofxz_surrogate_sources}, takes about 92.7~s, while subsequent evaluations for each new source location takes merely 0.11~s. On the contrary, it takes about 1.5~s for the fast sweeping solver to obtain a single solution. Therefore, the computational cost for both approaches would be the same for the first 67 source locations, after which each PINN solution is about 13.5 times faster. Given the fact that usually thousands of such source evaluations are needed for a given velocity model, the computational attractiveness of PINNs is obvious. These computations are done using an NVIDIA Tesla P100 GPU.

Moreover, the solution computed with the PINN surrogate model is considerably more accurate and, therefore, if we need similar accuracy with the fast sweeping method, we either need to use a much smaller grid spacing or a high-order version, leading to a substantial increase in the computational cost of the fast sweeping solver. Such high accuracy is often required for quantities that rely on traveltime derivatives, such as amplitudes and take-off angles. While these traveltime derivatives need to be separately computed for conventional methods, we obtain them as a by-product of the PINN training process.

Nevertheless, it is worth noting that the actual computational advantage of using the proposed algorithm, compared to conventional numerical solvers, depends on many factors, including the neural network architecture, optimization hyper-parameters, and sampling techniques. If the initialization of the network and the learning rate of the optimizer are chosen carefully, the training can be completed quite efficiently. Moreover, the type of activation function used, the adaptive weighting of the loss function terms, and the availability of second-order optimization techniques can accelerate the training significantly. Additional factors that would dictate the computational advantage include the complexity/size of the velocity model and the number of seismic sources. Therefore, a detailed study is needed to quantify the computational efficiency afforded by the proposed PINN eikonal solver compared to conventional algorithms by considering the afore-mentioned factors. Since the computational cost of solving the anisotropic eikonal equation is 3-5 times that of the isotropic case~\cite{waheed2015iterative}, PINNs will be computationally more attractive for traveltime modeling in anisotropic media.

Other advantages of the proposed framework include the ease of extension to complex eikonal equations, for example, the anisotropic eikonal equation, by simply modifying the neural network’s loss function. Since the method is mesh-free, contrary to conventional algorithms, incorporating topography does not require any special treatment. Furthermore, the point-source location does not need to be on a regular grid as needed by conventional numerical algorithms. This often results in using a smaller than necessary computational grid for conventional methods, thereby increasing the computational load or incurring errors by relocating the point-source to the nearest available grid point. Furthermore, our PINN eikonal solver uses \texttt{Tensorflow} at the backend, which allows easy deployment of computations across a variety of platforms (CPUs, GPUs) and architectures (desktops, clusters). On the contrary, significant effort needs to be spent on adapting the conventional algorithms to benefit from different computational platforms or architectures.

Nevertheless, there are a few challenges in our observation that need further investigation. Most importantly, we observe slow convergence for velocity models with sharp variations. Although the use of a second-order optimization method, locally adaptive activation function, and adaptive weighting of the terms in the loss function considerably improves the convergence rate, additional advances are needed, particularly to make PINNs computationally feasible for traveltime modeling of areas with complex subsurface geologic structures. Additional solutions may include a denser sampling of training points around parts of the velocity model with a large velocity gradient~\cite{anitescu2019artificial} or the use of non-local PINN framework~\cite{haghighat2020nonlocal}.  Another challenge is concerning the optimal choice of the network hyper-parameters that are often highly problem-dependent. Recent advances in the field of meta-learning may enable the automated selection of optimum architectures.

\section{Conclusions}

We proposed a novel algorithm to solve the eikonal equation using a deep learning framework. Through tests on benchmark synthetic models, we show that the accuracy of the proposed approach is better than the first-order fast sweeping solution. Depending on the heterogeneity in the velocity model, we also note that training is needed for only a fraction of the total grid points in the computational domain to reliably reconstruct the solution. We also observed that transfer learning could be used to speed up convergence for new velocity models and/or source locations. Moreover, having computed solutions corresponding to a few source locations for a given velocity model, surrogate modeling can be used to train a network to instantly yield traveltime solutions corresponding to new source locations. These properties, not afforded by the conventional numerical algorithms, potentially allow us to massively speed up seismic inversion applications, particularly for large 3D models. Moreover, the extension of the proposed framework to more complex eikonal equations, such as the anisotropic eikonal equation, requires only an update to the loss function according to the underlying PDE. Furthermore, contrary to conventional methods, the mesh-free nature of the proposed method allows easy incorporation of surface topography, which is often an important consideration for land seismic data.

We must, however, note that the conventional algorithms to solve the eikonal equation have evolved over several decades, gradually improving in their performance. While we have demonstrated the immense potential of the proposed framework, further study is needed to meet the robustness and efficiency required in practice. We demonstrated the proposed framework on 2D models for the simplicity of illustration. The extension to 3D velocity models is straightforward. 

\section{Acknowledgments}

We extend gratitude to Prof. Sjoerd de Ridder and three anonymous reviewers for their constructive feedback that helped us in improving the paper.
\\

{\noindent \bf Computer code availability}

All accompanying codes are publicly available
at \url{https://github.com/umairbinwaheed/PINNeikonal}.

\printcredits

\bibliographystyle{cas-model2-names}

\bibliography{pinn_eikonal}

\end{document}